\documentclass[reprint,dvipsnames,superscriptaddress,amsmath,amssymb,prx]{revtex4-2}
\usepackage{graphicx}           
\usepackage{dcolumn}            
\usepackage{bm}                 
\usepackage{physics}            
\usepackage{mathtools}          
\usepackage{nicefrac}           
\usepackage[usestackEOL]{stackengine} 

\usepackage[dvipsnames]{xcolor} 
\usepackage{hyperref}           
    \hypersetup
    {   
        unicode=true,           
        pdftoolbar=true,        
        pdfmenubar=true,        
        pdffitwindow=false,     
        pdfstartview={FitH},    
        pdfauthor={Gabriele Bellomia, Carlos Mejuto-Zaera, Massimo Capone, Adriano Amaricci},
        pdftitle={Quasilocal entanglement across the Mott-Hubbard transition},
        pdfkeywords={metal-insulator transition, entanglement measures, quantum information theory, quantum correlations in quantum information, cluster methods, dynamical mean-field theory, exact diagonalization, configuration interaction, Hubbard model, strongly correlated systems},
        pdfnewwindow=true,      
        colorlinks=true,        
        linkcolor=MidnightBlue, 
        citecolor=Maroon,       
        filecolor=Brown,        
        urlcolor=MidnightBlue   
    }

\newcommand{\rife}[1]
{\cite{#1}}

\newcommand{\figu}[1]
{Fig.\,\ref{#1}}

\newcommand{\secu}[1]
{Sec.\,\ref{#1}}

\def\LL{{\cal L}}

\def\eg{\mbox{\it e.g.\ }}
\def\ie{\mbox{\it i.e.\ }}

\def\up{\uparrow}

\def\dw{\downarrow}

\def \nnS#1{s_{\expval{#1}} }
\def \nnI#1{I_{\expval{#1}} }
\def \nnE#1{E_{\expval{#1}} }

\begin{document}

\title{Quasilocal entanglement across the Mott-Hubbard transition}

\author{Gabriele Bellomia}
    \email{gabriele.bellomia@sissa.it}
    \affiliation{International School for Advanced Studies (SISSA), Via Bonomea 265, 34136 Trieste, Italy}

\author{Carlos Mejuto-Zaera}
    \email{cmejutoz@sissa.it}
    \affiliation{International School for Advanced Studies (SISSA), Via Bonomea 265, 34136 Trieste, Italy}

\author{Massimo Capone}
    \email{capone@sissa.it}
    \affiliation{International School for Advanced Studies (SISSA), Via Bonomea 265, 34136 Trieste, Italy}
    \affiliation{Istituto Officina dei Materiali (CNR-IOM), Via Bonomea 265, 34136 Trieste, Italy}

\author{Adriano Amaricci}
    \email{amaricci@sissa.it}
    \affiliation{Istituto Officina dei Materiali (CNR-IOM), Via Bonomea 265, 34136 Trieste, Italy}   


\begin{abstract}

The possibility to directly measure, in a cold-atom quantum simulator, the von Neumann entropy and mutual information between a site and its environment opens new perspectives on the characterization of the Mott-Hubbard metal-insulator transition, in the framework of quantum information theory.
In this work we provide an alternative view of the Mott transition in the two-dimensional Hubbard model in terms of rigorous \emph{quasilocal} measures of entanglement and correlation between two spatially separated electronic orbitals, with no contribution from their environment.
A space-resolved analysis of cluster dynamical mean-field theory results elucidates the prominent role of the \emph{nearest-neighbor} 
entanglement in probing Mott localization: both its lower and upper bounds sharply
increase at the metal-insulator transition.
The two-site entanglement \emph{beyond
nearest neighbors} is shown to be quickly damped as the
inter-site distance is increased.
These results ultimately resolve a conundrum of previous analyses based on the single-site von Neumann entropy, which has been found to monotonically decrease when the interaction is increased.
The quasilocal two-site entanglement recovers instead the distinctive character of Mott insulators
as strongly correlated quantum states, demonstrating its central role in the $2d$ Hubbard model.
\end{abstract}

\maketitle

\section{Introduction}\label{intro}
The Hubbard model is a cornerstone of condensed matter physics as the
paradigmatic description of systems with strongly correlated electrons
and, in its two-dimensional version, as a central building block to
understand high-temperature superconductivity in copper
oxides. Besides the investigation of the various landmarks of the
cuprate phase diagram,  like $d$-wave superconductivity, charge-ordering/stripes and pseudogap,
a prominent role is played by the Mott metal-insulator transition (MIT),
which is arguably the most direct signature of strong correlations \cite{Mott_RMP, MIT_RMP}.

On the other hand, the flourishing of quantum  information theory has
greatly emphasized the possibility to quantify  the entanglement of
quantum many-body systems and its central role in the study of quantum
phase transitions
\cite{Osterloh_Nature,QuasiocalEntanglement_simpleQPT,Vidal_Entanglement_in_QCP,squillante2023gruneisen},
topological order \cite{Kitaev_TopologicalEntropy,Levin_TopologicalOrder,Wen_TopologicalOrder},
chemical bonding
\cite{Entanglement_and_Bonding,Correlation_and_bonding,Schilling_QuantumScience}
and the development of highly optimized numerical methods
\cite{White_DMRG,Optimizing_DMRG,Automated_Active_Spaces}.

In the last few years we have seen the first efforts to revisit
traditionally successful descriptions of the  Mott-Hubbard transition
in terms of this new  approach.
Initial studies within single-site dynamical mean-field
theory (DMFT) \cite{DMFT_RMP} have attempted to probe the development
of entanglement across the Mott transition \cite{XiDai2013}. The
spatially local nature of the correlations in this framework makes it natural to use
on-site markers like \emph{single-site entanglement entropy}
or \emph{single-site mutual information}. 

A similar approach has been more recently
pursued \cite{WalshLetter2019,WalshPRXQuantum2020,WalshPNAS2021}
within cluster dynamical mean-field theory (CDMFT), which includes the
effects of short-range spatial correlations within finite
clusters~\cite{Kotliar2001,ClusterRMP2005,Haule2008} onto these single-site
entropic measures.

The choice of these markers is driven by
their direct experimental access in cold-atom experiments
\cite{Cocchi_ColdAtoms} but it is also justified \emph{a priori} by the expectation that,
at least in a paramagnetic Mott insulator without long-range order, the 
particles are essentially localized and the relevant correlation functions
are expected to be short ranged.
Yet, the evolution of the single-site entanglement entropy as a
function of the interaction strength does not seem to
convey the expected physical picture for a correlation-driven localization.

In fact, despite being clearly influenced by the
Mott transition, this quantity decreases from the metallic to the insulating
phase~\cite{XiDai2013,WalshLetter2019}.
This seems in sharp conflict with the intuitive notion that nearly localized Mott insulators
are underpinned by the development of some kind of entanglement among electronic 
orbitals at different sites.
We can connect this shortcoming with the fact that, regardless of the
approximation, the single-site entanglement entropy in the half-filled 
one-band Hubbard model closely follows the behavior of the local double
occupancy, which naturally decreases across the Mott
transition \cite{XiDai2013}.
At a more fundamental level, we notice that the single-site entanglement
entropy results from contributions of any kind of nonlocal quantum correlations (bi/multipartite terms at 
different spatial ranges), and it does not directly
reflect the specific behavior of any such term \cite{EntanglementRMP}.
Remarkably, recent extensions of the analysis 
to multisite entanglement entropies \cite{Renyi_DMFT,Renyi_Honeycomb} 
have established the same qualitative picture,
suggesting that nonlocal entanglement needs to
be addressed carefully to uncover its role in
Mott localization.

In this work we overcome these limitations by considering \emph{quasilocal}
measures of entanglement, namely the entanglement between two lattice sites 
at distances ranging from nearest-neighbors, playing a major role, to more distant 
pairs. 
We compute these objects within zero-temperature CDMFT,
using a combination of numerical methods:
a Lanczos/Arnoldi exact diagonalization scheme~\cite{DMFT/ED_Capone,EDIpack} and the recently
introduced configuration-interaction based solver~\cite{ASCIdmft}, 
which gives access to the larger clusters required to assess 
the entanglement beyond nearest neighbors. 
In order to identify the fundamental information related to the Mott
transition we consider paramagnetic solutions in which magnetic
ordering is inhibited, as it has been done also in the previous
works where the connection between the MIT and entanglement
estimators has been addressed~\cite{XiDai2013,WalshLetter2019,WalshPRXQuantum2020,WalshPNAS2021}.

We will adopt and discuss different quantities, focusing on the
estimation of the orbital entanglement between lattice sites. 
We compute upper and lower bounds for the two-site
entanglement, by using suitable correlation measures~\cite{Wolf,Schilling_QuantumScience}
and physically motivated superselection rules on the relevant reduced density matrices~\cite{Schilling_Entanglement_and_Locality_2023,ding2023physical_entanglement}.
This allows us to obtain a terse physical picture of quantum and classical correlations
across the Mott transition in the $2d$ Hubbard model, as well as a
distinctive characterization of the metallic and insulating phases.
In particular, we show that the nearest-neighbor entanglement of the
Mott insulator is larger with respect to the Fermi liquid state,
recovering the intuitive expectations for a strongly correlated electronic system.\\

The rest of this work is organized as follows.
In \secu{entanglement_measures} we will present the  
measures of quasilocal correlation and entanglement. We further discuss
the selected superselection rules and elucidate their role
in the construction of lower bounds to the two-site 
entanglement.  
In \secu{model_methods} we briefly introduce the model and the
CDMFT solvers used to approximate its solution.
In addition we discuss the evaluation of
single- and two-site reduced density matrices within the chosen
numerical schemes. 
In \secu{all_results} we present and discuss our results
about the evolution of correlations and entanglement across the Mott
transition. Section \ref{SSR_relevance} contains a discussion on the
robustness of the single- and two-site measures of entanglement with respect
to the superselection rules, in the two phases of the model. 
In \secu{conclusions} we draw conclusions and outline some future perspectives.
Finally, in \appendixname~\ref{appendix:schillings_formulae} we give further details on the implementation of the superselected measures of correlation and entanglement,
in \appendixname~\ref{appendix:subadditivity} we outline the
derivation of a sum rule for the two-site mutual information,
in \appendixname~\ref{appendix:fordummies} we summarize in a
table the properties of the entanglement and correlation 
markers used in this work
and in \appendixname~\ref{appendix:ASCI} we provide additional
details on the numerical calculations.

\section{Entanglement measures}\label{entanglement_measures}
The key quantity for our analysis is the quasilocal orbital entanglement,
as measured by the relative entropy of entanglement (REE)~\cite{Vedral_RMP}.
The relationships between the REE and different measures of entanglement,
\eg entanglement of formation, entanglement of distillation, etc.
are extensively discussed in Refs.\,\cite{Vedral_RMP,Schilling_QuantumScience}.

In the following, we mostly deal with the case of bipartite entanglement between 
two spatially separated electronic orbitals.  
Thus, in order to formally define the corresponding REE, we consider two sites, not necessarily neighbors, as a
subsystem of a full lattice.
This subsystem can be further partitioned into the two
distinct sites. The REE between these two sites corresponds to a
minimal ``distance'' (quantum relative entropy) between the given two-site density matrix and the convex
set of two-site states that are separable with respect to the bipartition~\cite{Vedral_RMP}.
In the following we will refer to this quantity as the two-site entanglement $E_{ij}$.
Since there is not a simple parametrization for this convex set, a
closed expression for the $E_{ij}$ remains elusive~\cite{krueger2005open}.
Yet, different bounds can be constructed via conventional quantities~\cite{Wolf,Schilling_QuantumScience}
and suitable superselection rules~\cite{Schilling_Entanglement_and_Locality_2023,ding2023physical_entanglement}.
In the following we describe the evaluation of these quantities in
terms of the ground state reduced density matrices for single- and
two-site subsystems, as computed or measured within any theoretical or
experimental method. We defer to \secu{model_methods} all the details 
specific to our numerical approach.

\subsection{Single- and two-site entanglement entropies}\label{EEntropy}
Let us consider the ground state density matrix
$\rho_\mathrm{gs}$ for a lattice $\LL$. The reduced density matrix (RDM) for a subsystem
$\mathrm{A}$ of the lattice $\LL$ reads 
\begin{equation}
    \rho_\mathrm{A} = \Trace_{j\in\LL\setminus\mathrm{A}}\left(\rho_\mathrm{gs}\right)\,.
  \end{equation}
Given a single site $i$, we call local entanglement entropy the von Neumann entropy of $\rho_i$~\cite{EntanglementRMP,XiDai2013,WalshLetter2019,WalshPRXQuantum2020,WalshPNAS2021}
\begin{equation}
    s_i = -\Trace(\rho_i\log\rho_i)
    \label{eq:local_entropy}
\end{equation}

If $\rho_\mathrm{gs}$ describes a pure state then $s_i$ gives a
well-defined measure of the bipartite entanglement between the
single site and the surrounding lattice~\cite{EntanglementRMP}. 
We stress that being one of the 
entangled parties a macroscopic subsystem, this quantity 
includes quantum correlations for all the spatial ranges 
encoded in the model, hence the name ``local entanglement
entropy'' can be misleading. 

In a similar way, we define a two-site entanglement entropy 
from the RDM of a pair of sites $\rho_{ij}$ as
\begin{equation}
  s_{ij} = -\Trace(\rho_{ij}\log\rho_{ij}).
  \label{eq:quasilocal_entropy}
\end{equation}
We indicate the particular case of nearest neighbors with the
symbol $\nnS{ij}$.   
The two-site entanglement entropy $s_{ij}$ shares most
properties with $s_i$, being it a measure of the entanglement 
between a pair of sites and, again, their whole environment.

\subsection{Two-site mutual information (total correlation)}\label{PairTotalCorrelation}
Given the entanglement entropies $s_i$ and $s_{ij}$, respectively 
of a single- and a two-site subsystem, we can define the mutual information
between the two sites as~\cite{EntanglementRMP}
\begin{equation}
    I_{ij} = s_i + s_j - s_{ij}\,,
    \label{eq:bond_correlation}
\end{equation}
reserving the symbol $\nnI{ij}$ for the case of nearest neighbors. \\

The mutual information $I_{ij}$ gives a measure of all quantum and classical 
correlations between sites $i$ and $j$.
To elucidate this, let us consider the density matrix of a generic 
bipartite quantum system $\rho_\mathrm{AB}$, with $\rho_\mathrm{A}$
and $\rho_\mathrm{B}$  the reduced density matrices for an arbitrary
pair of subsystems.
The mutual information $I_\mathrm{AB} = s_\mathrm{A} + s_\mathrm{B} -
s_\mathrm{AB}$, satisfies the following inequality \cite{Wolf}
\begin{equation}
    I_\mathrm{AB} \geq \frac{\left(\expval{\mathcal{O}_\mathrm{A} \otimes \mathcal{O}_\mathrm{B}}_{\rho_\mathrm{AB}} - \expval{\mathcal{O}_\mathrm{A}}_{\rho_\mathrm{A}}\expval{\mathcal{O}_\mathrm{B}}_{\rho_\mathrm{B}}\right)^2}{2\norm{\mathcal{O}_\mathrm{A}}^2\norm{\mathcal{O}_\mathrm{B}}^2}
    \label{eq:total_correlation}
\end{equation}
where $\mathcal{O}_\mathrm{A}$ and $\mathcal{O}_\mathrm{B}$ are generic
operators acting on the Hilbert spaces of subsystems A and B, respectively.
If the composite system is pure
($s_{\mathrm{AB}}=0$), we have $I_{\mathrm{AB}} = s_\mathrm{A} + s_\mathrm{B} = 2s_\mathrm{A}$. 
Because there is no classical correlation in a pure state, the mutual information between
A and B measures only the quantum correlation between the two.
If $s_{\mathrm{AB}}\neq0$ then $\rho_{\mathrm{AB}}$ is
a statistical mixture and $I_{\mathrm{AB}}$ also includes 
classical correlations between A and B.
Hence, $I_\mathrm{AB}$ provides an upper bound to correlation functions 
for any pair of subsystem operators, namely it quantifies the maximum 
correlation between subsystems A and B.

\subsection{Entanglement bounds and superselection rules}\label{SSRules}
Since a closed mathematical expression for the two-site entanglement,
as measured by the REE, is inaccessible we aim at obtaining upper and lower bounds
to $E_{ij}$. As for the two-site entanglement entropy and mutual information, we
will notate the case of nearest neighbors with the symbol $\nnE{ij}$.

As discussed above, the total correlation $I_{ij}$ includes different
sources of correlations. Hence it provides a straightforward upper bound to the
two-site entanglement $E_{ij}\leq I_{ij}$ \cite{Schilling_QuantumScience,Vedral_RMP,EntanglementRMP}.
A lower bound can be obtained by restricting the family of subsystem
density matrices over which the distance in the definition of the REE
is evaluated~\cite{Schilling_Entanglement_and_Locality_2023}.
This restriction can be motivated by assuming suitable \emph{local} superselection rules (SSR),
which are generalizations of conventional selection rules, constraining the coherent superposition of states pertaining to different eigenvalues of selected local operators~\cite{Schilling_Entanglement_and_Locality_2023,ding2023physical_entanglement}. 
The relevance of local SSR in the definition of physically accessible
entanglement~\cite{Wick1952,Wick1970,Bartlett2003,Banhuls2007,Friis2013,Friis2016,Benatti2020,Vidal2021,RicardoCosta2021,ding2023physical_entanglement,DelMaestro_SSR}
has been recently challenged by new arguments in quantum
thermodynamics \cite{Esposito2023}.
Notwithstanding their possible interpretation, it has been shown that
the SSR allow the definition of rigorous lower bounds to the two-site REE~\cite{Schilling_Entanglement_and_Locality_2023}.

In the following, we consider particle number (N-SSR) and parity
(P-SSR) superselection rules, corresponding to a restriction on the 
physical operations
allowed on the electronic system, namely they must conserve either the
local number of electrons $N_i$ or its parity. Using these SSR we
obtain the following bounds for the two-site entanglement:
\begin{equation}
    E_{ij}^\text{N-SSR} \leq E_{ij}^\text{P-SSR} \leq E_{ij}\,.
\end{equation}
Explicit expressions for $E_{ij}^\text{N-SSR}$ and
$E_{ij}^\text{P-SSR}$ in terms of the components of the two-site
reduced density matrices are reported in
\appendixname~\ref{appendix:schillings_formulae}.

It is important to observe that the N-SSR and P-SSR can
also be applied to the local entanglement and the two-site
total correlation, providing lower bounds to their magnitudes~\cite{EntanglementVsCorrelation2021}
\begin{align}
    E_i^\text{N-SSR} \leq E_i^\text{P-SSR} &\leq E_i \\[0.5ex]
    I_{ij}^\text{N-SSR} \leq \,I_{ij}^\text{P-SSR} &\leq I_{ij}
\end{align}
A self-contained presentation of explicit expressions for these
quantities is given in \appendixname~\ref{appendix:schillings_formulae}. 
Here, we denote with $E_i^\text{N(P)-SSR}$ the superselected measure of
the entanglement between a single site and the rest of the
lattice. Note that this quantity
does not in general correspond to the superselected local entanglement entropy
as the ground state density matrix of the lattice cannot be guaranteed to be pure under
the SSR~\cite{EntanglementVsCorrelation2021}, hence 
$$E_i^\text{N(P)-SSR} \neq s_i^\text{N(P)-SSR}.$$
See \appendixname~\ref{appendix:schillings_formulae} for an extended discussion.\\

Finally, we introduce the \emph{superselection factors}
\begin{align}
  \xi^\text{N(P)-SSR}_{E_i}&\overset{\text{def}}{=}\frac{E_i}{E_i^\text{N(P)-SSR}} \geq 1 \label{eq:local_ssr_factor}\\[0.5ex]
  \xi^\text{N(P)-SSR}_{I_{ij}}&\overset{\text{def}}{=}\frac{I_{{ij}}}{I_{{ij}}^\text{N(P)-SSR}} \geq 1 \label{eq:nn_ssr_factor}
\end{align}
in order to quantify the robustness of these entanglement and correlation measures, with respect to the two selected local
superselection rules: the closer these factors are to the unity, the less relevant the superselection rules become in defining the
corresponding measure.

\section{Model and methods}\label{model_methods}
We consider a single-band Hubbard  model on the two-dimensional square
lattice $\LL$, with Hamiltonian
\begin{equation}
H = -t\sum_{\sigma}\sum_{\expval{ij}\in\LL} \left(c^\dagger_{i\sigma}c_{j\sigma} +
  \mathrm{h.c.}\right) + U\sum_{i\in\LL} n_{i\up}n_{i\dw}
\end{equation}
where $c_{i\sigma}$
($c^\dagger_{i\sigma}$) is the annihilation (creation) operator of an
electron with spin $\sigma$ at the site $i$ of the lattice,
$n_{i\sigma}=c^\dagger_{i\sigma}c_{i\sigma}$ is the local spin density,  
$t$ is the nearest-neighbor hopping amplitude and $U$ is the local
electronic interaction strength.
In the following, we consider the
regime of zero temperature and an occupation of one electron per site (half-filling).
We set our unit of energy as the noninteracting 
half-bandwidth $D=4t=1$. 

We solve this model using cluster dynamical mean-field theory
(CDMFT)~\cite{DMFT_RMP,Kotliar2001,ClusterRMP2005,Haule2008}. This is a powerful 
nonperturbative tool to investigate the physics of
strongly correlated electrons on a lattice, capturing local as
well as nonlocal (short-range) correlations.
The CDMFT solution maps the original lattice problem to
a quasilocal embedding of a cluster of $N_\mathrm{imp}$ correlated sites
into a noninteracting bath, which is self-consistently determined by the
requirement that the Green's functions connecting sites within the cluster are the same as in
the original lattice. This effective theory can be seen as a generalization of a quantum impurity model, 
hence we will refer to the $N_\mathrm{imp}$ cluster sites as impurity sites in the following.

In this work, we rely on a combined numerical analysis using two
different methods to solve the quantum impurity cluster problem,
namely Lanczos/Arnoldi exact diagonalization (ED) and adaptive sampling
configuration interaction (ASCI).
Both methods parametrize the noninteracting bath in
terms of a finite number of sites, employing
different algorithms to obtain the lower part of the energy spectrum
and dynamical correlation functions.
Given the unfavorable scaling of the computational effort with the number of impurity sites, the size of the clusters is 
severely limited allowing for a reliable estimate only of short-ranged spatial correlations.

Our ED solver leverages on the massively parallel implementation of Ref.\,\cite{EDIpack}, straightforwardly generalized to cluster methods~\footnote{An unpublished version of the implemented cluster extension can be found at \url{https://doi.org/10.5281/zenodo.10628157}}. To reduce the number of variables to be optimized, we employ a representation of the 
bath in terms of non-interacting \emph{replicas} of the correlated cluster, diagonally coupled to the 
corresponding impurity sites~\cite{Capone2004PRB,Koch2008PRB}. Their internal parameters,
corresponding to on-site energies and nearest-neighbor hoppings are
optimized through the self-consistency procedure, together with the
impurity-bath hybridization amplitudes~\cite{EDIpack}.
Thus a given number $N_\mathrm{repl}$ of replicas
corresponds to $N_\mathrm{bath}=N_\mathrm{imp}N_\mathrm{repl}$ bath sites. 
In the following, we use ED to study
$N_\mathrm{imp}=1\times2$ and $N_\mathrm{imp}= 2\times2$ clusters,
keeping a fixed total number of sites $N_\mathrm{s} = N_\mathrm{imp}+N_\mathrm{bath} =
N_\mathrm{imp}(1+N_\mathrm{repl}) = 12$. 

In addition we employ the recently introduced ASCI solver for
CDMFT~\cite{ASCIdmft,MejutoZaera2021,Levine2020,Mejuto2022} to
benchmark our results and study larger cluster sizes with respect to
those accessible with the ED algorithm.
This method provides a powerful
algorithm~\cite{Tubman2016,WilliamsYoung2023} capable of
alleviating the ED limitations related to
the exponential growth of the Hilbert space of the impurity problem:
the ASCI Ansatz corresponds to an adaptively optimized truncation of
the full impurity model Hilbert space in terms of a subset of selected
Slater determinants, which together reconstruct the majority of the ground state wave
function. It further relies on optimizing an orbital active space by constructing a suitable natural basis from an approximated one-body RDM for the
bath orbitals~\cite{ASCIdmft,MejutoZaera2021,Mejuto2022}.   
This results in a highly compact and accurate representation of the ground state~\cite{Bravyi2016} and
of the one-body Green's functions~\cite{ASCIdmft}, while reducing the
overall computational cost.

Within the ASCI method the bath sites are split into groups of $N_\mathrm{imp}$
degenerate levels, with an all-to-all amplitude to the cluster
impurity sites. The bath parameters are self-consistently determined
by means of a recently introduced conic optimizer~\cite{ASCIfit}.
In this work we perform ASCI calculations addressing $N\times2$
clusters sizes, with $N=1,\dots,4$. We fix the number of bath sites to six
times the number of cluster impurity levels which, using the bath degeneracy
structure in Ref.\,\cite{ASCIfit}, means that we have six nondegenerate
bath energies.

The two methods, ED and ASCI, grant us direct access to 
an explicit representation of the ground state of the
cluster impurity model. Using this information, we can build the 
zero temperature RDM for any cluster subsystem 
(\eg two sites) using an on-the-fly trace over the bath and 
complementary impurity degrees of freedom.
The local RDM, thoroughly studied in
Refs.\,\cite{XiDai2013,WalshLetter2019,WalshPRXQuantum2020,WalshPNAS2021},  
is obtained by further tracing over all but one site.

Remarkably, we observe that a recent work by Roósz \emph{et
al.}\,\cite{Held_2RDMfrom2GF} provides a recipe to obtain
single-orbital and two-orbital RDM from the knowledge of
single-particle and two-particle Green's functions alone,
giving access to quasilocal entanglement and correlation
measures to a broad multitude of many-body methods.
Finally, an experimentally feasible tomography protocol has been recently proposed for dot-cavity devices, giving access to the relevant single-site and two-site RDM \cite{LidiaStocker2022}. Hence we prospect cross-fertilization
with the emerging field of entanglement certification in realistic open quantum impurity systems \cite{Carisch2023}.

\section{Results}\label{all_results}

In this section we present numerical results for the evolution of
quasilocal entanglement and correlations, as functions of the ratio $U/D$ across the interaction driven metal-insulator transition.
The critical interaction for the transition depends on the size of the cluster $N_\mathrm{imp}$.  We tracked the transition
point for all the investigated cases and estimated it to be placed in
the interval $U/D=[1.5,1.6]$, although a precise determination of the
transition point is beyond the scope of this work. 

\subsection{Mott transition in the entanglement entropy}\label{MIT_Local_Pair}

\begin{figure*}
  \includegraphics[width=0.46\textwidth]{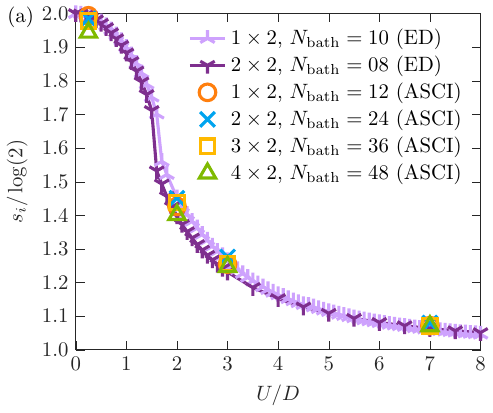}
  \hspace{1cm}
  \includegraphics[width=0.46\textwidth]{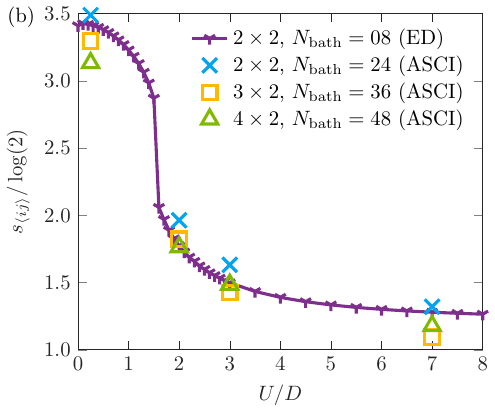}
  \caption{
    Local (a) and nearest-neighbor (b) entanglement entropies,
    respectively $s_i$  and $\nnS{ij}$, as a
    function of the interaction strength $U$ across the
    metal-insulator transition. 
    Data from ED (lines and symbols) and ASCI
    (open symbols) and for different combinations of cluster
    sizes and bath levels. Clusters have the shape  $N\times2$. 
  }      
  \label{fig:local_scaling}
\end{figure*}

We begin by investigating the local and the nearest-neighbor
entanglement entropies across the Mott transition.
In \figu{fig:local_scaling}
we report the behavior of  $s_i$ and
$\nnS{ij}$ as a function of the interaction strength. 
We compare the results for different cluster sizes and effective
bath realizations, from both ED and ASCI methods.  

The local entanglement entropy, see \figu{fig:local_scaling}(a), is
essentially insensitive to the cluster size, even in the
proximity of the transition point at intermediate interaction values. 
This confirms that the entanglement between a single-site and its environment
is properly captured in CDMFT and that this quantity, extensively treated
in the recent literature
\cite{WalshLetter2019,WalshPRXQuantum2020,WalshPNAS2021}, indeed represents a
suitable marker for the Mott transition.

The nearest-neighbor entanglement entropy $\nnS{ij}$, see \figu{fig:local_scaling}(b),
displays a more noticeable scaling with respect to the system size.
In particular, major quantitative discrepancies arise 
between clusters of different sizes in the metallic and insulating
regimes. We relate this behavior to the fact that some choices of the cluster
shape (\eg $3\times 2$ and $4\times 2$) can break the symmetries of the full lattice.  
However, according to their qualitative behavior across the Mott
transition we can interpret the nearest-neighbor entanglement entropy to be
well converged.

Having clarified the behavior of the local and nearest-neighbor entanglement entropies 
with respect to system size, in the following we shall focus on the
$2\times 2$ cluster for an in-depth analysis of the entanglement properties. 
A more thorough study of the scaling of the results beyond the nearest-neighbor limit in larger clusters will be investigated in \secu{BeyondNN} to further characterize
the nonlocality of classical and quantum correlations.

To start, we study the behavior of the entanglement entropies in the two
limiting regimes of interaction.  
In the noninteracting limit $U \ll D$ the local entanglement entropy approaches
the value $s_i=2\log(2)$, corresponding to a maximally entangled single-site
state, for a 
half-filled noninteracting metal featuring equal populations of
empty, doubly and singly occupied electronic states. 
On the contrary we have not a maximally entangled state for
nearest-neighbor sites, so the corresponding two-site entanglement entropy
attains a value smaller than $4\log(2)$. This is a direct consequence of the spatial correlation between the sites $i$ and $j$: if $s_{\expval{ij}}=4\log(2)$, then we would have $I_{\expval{ij}}=0$, which according to Eq.~\ref{eq:total_correlation} would imply that no finite correlator exist between the two neighboring sites.
Nevertheless, a spectral analysis of the nearest-neighbor RDM reveals that,
in this regime, all the 16 pure states contribute, although differently, to the configuration of the system, as expected for the noninteracting character of the solution.

In the strong coupling regime, $U\gg D$, both the local and
nearest-neighbor von
Neumann entropies appear to approach $\log(2)$.
Yet, a deeper analysis of the underlying pure states reveals
remarkable differences in the two corresponding density matrices.
The local entanglement entropy is largely dominated by the equally weighted 
two pure states $\ket{\uparrow}$ and $\ket{\downarrow}$, namely the
well-known local description of a paramagnetic Mott insulator.
On the other hand, $\nnS{ij}$ is dominated by four pure states, the spin-singlet
$(\ket{\uparrow\downarrow}-\ket{\downarrow\uparrow})/\sqrt2$,
accounting for about 75\% of the statistical mixture, and the
spin-triplet states $\ket{\uparrow\uparrow}$,
$\ket{\downarrow\downarrow}$ and
$(\ket{\uparrow\downarrow}+\ket{\downarrow\uparrow})/\sqrt2$, adding
up for almost all the rest, in equal parts.
Such relative composition of singlet and triplet states in
the nearest-neighbor dimer has been checked to be consistent across all
the addressed cluster and bath sizes: the numerical differences in the
von Neumann entropies for the larger clusters are due to a different
cumulative weight of all the remnant 12 pure states. 
This once again confirms that the two-site reduced density matrices are
qualitatively consistent across all the cluster realizations.

\subsection{Entanglement between nearest neighbors}\label{NNPairEntanglement}
\begin{figure}
    \includegraphics[width=0.42\textwidth]{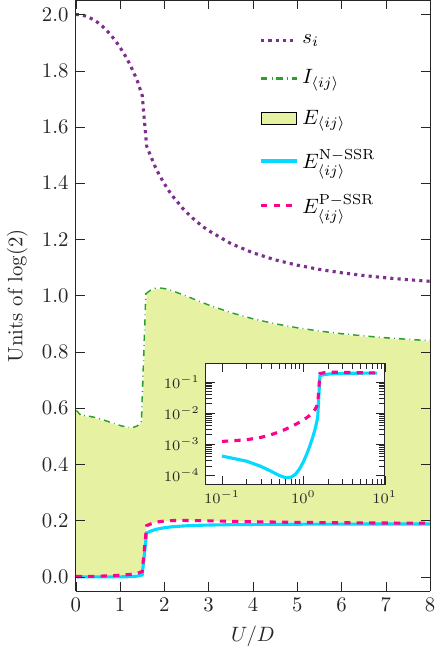}
    \caption{\label{fig:bond_entanglement}
      Local entanglement entropy $s_i$,
      nearest-neighbor total correlation $\nnI{ij}$ and the
      number (parity) superselected nearest-neighbor entanglement
      $\nnE{ij}^\text{N(P)-SSR}$ as a function of the interaction strength.
      The latter curves are also reported in logarithmic scale in the
      inset, highlighting the different behavior in the Fermi liquid.     
      The shaded area defines the range in which the unconstrained
      nearest-neighbor entanglement $\nnE{ij}$ lies. 
      Data from $2\times2$ cluster calculations using ED method with
      $N_\mathrm{bath}=8$. 
}
\end{figure}

Here, we investigate in more detail the quantum correlation 
between adjacent lattice sites using the $2\times2$ plaquette CDMFT/ED
solution of the model, which captures the essential features of the
local and nearest-neighbor density matrices. 
In \figu{fig:bond_entanglement} we show the nearest-neighbor total
correlation $\nnI{ij}$ and the particle number (parity) superselected
nearest-neighbor entanglement $\nnE{ij}^{\text{N(P)-SSR}}$, compared to
the well-known behavior of the local entanglement entropy $s_i$.
Considering the bounds  $\nnE{ij}^{\text{N(P)-SSR}} \leq \nnE{ij} \leq \nnI{ij}$,
we can identify a region (shaded area in \figu{fig:bond_entanglement}) 
in which the unconstrained nearest-neighbor entanglement lies.
Remarkably, both the upper and lower bounds to $\nnE{ij}$ show a sudden
rise at the Mott transition,
in sharp contrast with the local entanglement entropy which is monotonically
suppressed as the interaction is increased. 
This is a key result of this work that we can rationalize as follows.
The local entanglement entropy includes contributions for all spatial ranges \cite{EntanglementRMP}.
Conversely, $\nnE{ij}$ takes into
account only the entanglement across a single lattice bond. 
Thus, when looking at the spatial entanglement 
of the system through $s_i$, the Mott insulator might appear less correlated than a
Fermi liquid state. Instead, the nearest-neighbor entanglement undergoes an
abrupt boost at the transition, recovering the usual picture of Mott
insulators as locally strongly correlated phases of matter.
We interpret this result as a signature of the intimate relationship
between strong correlations and (Mott) localization. 

We can push the resulting physical picture further by focusing on the comparison between the local entanglement entropy and the nearest-neighbor total
correlation. Elaborating on the strong subadditivity property of the von Neumann entropies
\cite{Lieb_SSA} we can write the following inequality (see \appendixname~\ref{appendix:subadditivity})
\begin{equation}
       \bar{I}_{ij} \overset{\text{def}}{=} \frac{1}{\ell-1}
       \sum_{j\neq i} I_{ij} \leq 2s_i,
    \label{eq:worked_out_SSA}
\end{equation}
where $\ell$ is the number of sites in the cluster and we
recall that $I_{ij}$ represents the total correlation between two
impurity sites $i$ and $j$. 
Hence the local entanglement entropy bounds from above the
\emph{cluster-averaged} two-site total correlation $\bar{I}_{ij}$.
The expression for $\bar{I}_{ij}$ contains $\zeta$ identical terms, namely 
$\nnI{ij}$, with $\zeta$ the number of nearest neighbors of site $i$ in the cluster.
We can assume that the remaining terms in the expression
for $\bar{I}_{ij}$ decay with the intersite
distance~\cite{Wolf,Tajik2023}.
Their contribution is then at most a negative additive shift to the
  value of the local entanglement entropy. This explains the similar
  tail behavior of $s_i$ and $\nnI{ij}$ in the Mott regime and
  implies that the total correlation in the Mott phase is more local,
  namely has shorter range, than the quantum entanglement of a Fermi liquid state.
More evidence about the quasilocal nature of entanglement and correlations in 
Mott insulators will be given in \secu{BeyondNN}. 

Finally, we note that both $\nnE{ij}^\text{P-SSR}$ and $\nnE{ij}^\text{N-SSR}$ 
essentially vanish throughout the whole metallic phase and quickly saturate to 
their maximum value in the Mott insulator.
The presence of two sharply distinct entanglement scales clearly identifies
the two phases, suggesting that these superselected measures are able to 
capture the adiabatic connection to either the noninteracting and the
strong coupling limit, at all intermediate strongly correlated regimes.
It is worth remarking that a deeper inspection (inset in \figu{fig:bond_entanglement})
reveals a qualitatively different behavior of $\nnE{ij}^\text{P-SSR}$ and $\nnE{ij}^\text{N-SSR}$,
throughout the Fermi liquid.
After the Mott transition the superselection rules instead become
quickly indistinguishable on the nearest-neighbor entanglement.
This can be readily understood by looking at their expressions, see Eqs.~\ref{eq:Eij_N-SSR} and \ref{eq:Eij_P-SSR}:
the two quantities differ by a term depending on the populations
of doublons and holons, which asymptotically vanish in the Mott phase.

\subsection{Entanglement beyond nearest neighbors}\label{BeyondNN}

\begin{figure}
    \includegraphics[width=0.45\textwidth]{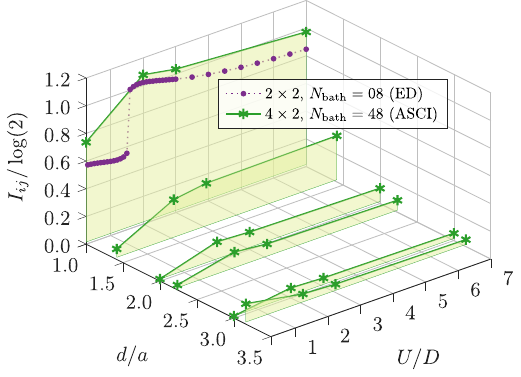}
    \caption{\label{fig:further_sites}
      Two-site total correlation $I_{ij}$ as a function of intersite distance $d/a$
      (with $a$ the lattice spacing) and interaction strength $U/D$.
      For $d/a=1$ it reduces to the nearest-neighbor total correlation $\nnI{ij}$
      (see main text).       
      Data from ASCI calculations using $4\times 2$ cluster size
      (star symbols and solid lines) and ED calculations for $d/a=1$ using
      $2\times2$ cluster (filled symbols and dotted line).     
      The shaded area represents $E_{{ij}} \leq I_{{ij}}$, namely the
      range in which lies the nonlocal entanglement between the two
      selected sites.      
}
\end{figure}

Having clarified the behavior of the two-site entanglement for the minimal lattice distance, we now devote our attention to its scaling beyond nearest neighbors.   
To this end we consider a $4\times2$ cluster which we address using
the ASCI method for CDMFT. 
In \figu{fig:further_sites} we show the total correlations 
$I_{ij}$ between two lattice sites $i$ and $j$, at increasing distances. 
In particular we choose the lower left corner site of the cluster 
as $i$, and vary $j$ as to select all the intracluster distances
$d = a, \sqrt{2}a, 2a, \sqrt{5}a, 3a, \sqrt{10}a$, where $a$ is the
lattice constant.
Given that $I_{ij}$ represents an upper bound for
the two-site entanglement, we can readily infer that the
two-site quantum correlations are quickly damped with distance.
Remarkably, the increase of two-site correlations at the Mott 
transition observed for nearest-neighbor
pairs is progressively smoothed at larger distances.
These findings further underline the quasilocal nature of the two-site entanglement across the whole phase diagram of the model, with the Mott insulator likely to have even a shorter range with respect to its parent Fermi liquid state.

\section{Robustness with respect to SSR}\label{SSR_relevance}
We sharpen our analysis by applying the N-SSR and the
P-SSR to the local entanglement entropy and the nearest-neighbor total 
correlation. This will enable us to understand how these quantities are
affected by the superselection rules, which are expected to give the hallmark of
experimentally accessible correlations in a realistic setup based on
operations performed onto individual local electronic degrees of freedom~\cite{Wick1952,Wick1970,Bartlett2003,Banhuls2007,Friis2013,Friis2016,Benatti2020,Vidal2021,RicardoCosta2021,ding2023physical_entanglement,Schilling_Entanglement_and_Locality_2023}.

\begin{figure}
    \includegraphics[width=0.45\textwidth]{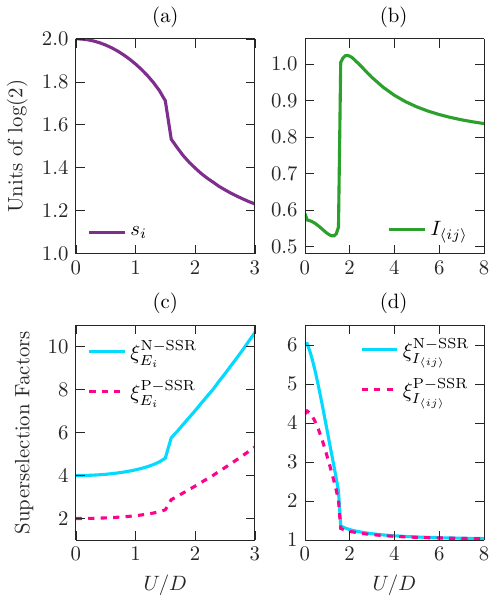}
    \caption{\label{fig:superselection}
      Local entanglement entropy (a), nearest-neighbor total correlation (b)
      and their respective [(c) and (d)] particle number and parity superselection
      factors $\xi^\text{N(P)-SSR}_{E_i}$ and $\xi^\text{N(P)-SSR}_{\nnI{ij}}$, as a
      function of the interaction strength.
      Data from ED calculations with a $2\times2$ cluster and
      $N_\mathrm{bath}=8$ bath levels.
    }
\end{figure}

In \figu{fig:superselection}, we report the local entanglement entropy
$s_i$ and the nearest-neighbor total correlation 
$\nnI{ij}$ together with the respective superselection factors, as defined in
Eqs.~\ref{eq:local_ssr_factor}, \ref{eq:nn_ssr_factor}.
The explicit expressions of these two quantities  
can be retrieved from \appendixname~\ref{appendix:schillings_formulae}, see Eqs.~\ref{eq:Ei_N-SSR}, \ref{eq:Ei_P-SSR} and Eqs.~\ref{eq:Iij_N-SSR}, \ref{eq:Iij_P-SSR} respectively.
These factors quantify the weight that the superselection rules
filter out from the corresponding correlation measures.

In the Fermi liquid both factors are strongly  enhanced by the SSR,
signaling that correlations and entanglement in
a metal are not really accessible via standard local operations.
This matches with the results previously discussed in \figu{fig:bond_entanglement}:
the most localized bipartite entanglement measure we can define on a
lattice vanishes in an interacting metal, provided that we preclude
any local fluctuation of charge. The local entanglement and two-site correlations, measured by $s_i$ and $I_{\expval{ij}}$, are then ascribed to the itinerant nature of the electronic state.

Remarkably, in the correlation-driven insulator we observe a dramatic
differentiation between the two markers. The single-site entanglement
entropy superselection factors $\xi^\text{N(P)-SSR}_{E_i}$ linearly increase 
with the on-site repulsion.  
On the contrary,
$\xi^\text{N(P)-SSR}_{\nnI{ij}}$ steadily approach the unity right after the
transition point, signalling that the results are increasingly robust with respect to
the SSR, \ie the filtering is nearly irrelevant in this regime. 

The fact that $s_i$ is damped by superselection rules while the 
nearest neighbor total correlation becomes essentially insensitive to
them, as we enter the Mott phase, suggests that nearest-neighbor
quantities are better suited for the characterization of correlated
insulators.
Indeed, the signature of Mott physics is the freezing of local charge 
fluctuations thus the N-SSR would constitute a natural physical 
stage to reveal the underpinning role of strong correlations.

The robustness of the nearest-neighbor total correlation to N-SSR
and P-SSR and the fact that  $E_{\expval{ij}}^\text{N-SSR} \simeq
E_{\expval{ij}}^\text{P-SSR}$ deep in the Mott insulating phase, suggests that the
nearest-neighbor entanglement $\nnE{ij}\leq \nnI{ij}$ could share the
same property with respect to the SSR. In this
case, we expect the full unrestricted quantity to be significantly
close to its lower bound shown in \figu{fig:bond_entanglement}.

\section{Conclusions and outlook}\label{conclusions}
The local entanglement entropy has been extensively studied through
the last decade to revisit the physics of the Mott transition under
the lens of quantum information theory
\cite{XiDai2013,WalshLetter2019,WalshPRXQuantum2020,WalshPNAS2021}.
Although this quantity allows for a full characterization of the
quantum phase transition and both its sub- and super-critical thermal
signatures, its physical interpretation has been hindered by
the lack of a clear distinction between genuinely local and nonlocal contributions~\rife{EntanglementRMP}. 

Based on a CDMFT analysis of the two dimensional Hubbard model
we clarified the role of local
and \emph{quasilocal} entanglement across the Mott transition. 
In particular, we leveraged a notion of entanglement between a pair of sites, 
with no reference to their environment.
This quantity gives a genuine measure of the quasilocal
entanglement in the ground state. Despite being well defined as the relative entropy
of entanglement, such two-site entanglement has no accessible general expression and its determination represents an open problem in quantum information theory \cite{krueger2005open}.
Consequently, we carefully analyze upper and lower bounds to the two-site
entanglement. The former is obtained by a rigorous interpretation of the 
mutual information as a measure of maximal total correlations, while the 
latter are based on recently introduced expression for the two-site entanglement
under charge and parity superselection rules~\cite{Schilling_Entanglement_and_Locality_2023}, amounting to constrain the possible coherent superposition of quantum states to those conserving the local electron number or its parity.

We proved that the nearest-neighbor total
correlation is almost unaffected  by the local charge and
parity superselection rules, in the Mott insulator. On the other hand the charge and parity superselected two-site entanglement formulas have been proven indistinguishable in the Mott phase. Hence we propose the two-site measures of entanglement and correlation as reliable tools for the study of (Mott) localized phases of matter.

Following the evolution of the entanglement bounds as a function of
the interaction strength, we predict a sharp increase of the nearest-neighbor entanglement at the Mott transition point, in contrast with the well-known phenomenology of the local entanglement entropy.
Consequently, while the Mott insulator might globally result less
spatially entangled than the weak coupling Fermi liquid state,  
we argue that the genuine quasilocal quantum correlation is actually
increased by Mott localization, thus reconciling with the paradigmatic 
view of a Mott insulator as a strongly correlated localized system.
Further evidence about the quasilocal nature of the two-site
entanglement, in both the Fermi liquid and Mott regimes, has been
secured by extending our analysis beyond nearest neighbors.

Our results shed new light on the mechanism underlying the 
transformation of a Fermi liquid metal into a strongly correlated insulator,
namely the Mott transition, bridging the fertile field of quantum information
theory with the notoriously tough problem of describing strongly correlated
electrons.
The analysis is based on CDMFT calculations with clusters sizes 
up to eight sites, providing a reasonably complete description of 
the Mott transition, as extensively studied and compared with
other methods, including larger clusters
\cite{Schafer21Multi,Schafer24Mott}.
The addressed cluster sizes allow us for a systematic and
computationally affordable study in a well-documented setting.
There are several
directions to verify the robustness of our results, including the
analysis of larger clusters within CDMFT or dynamical cluster
approximations \cite{ClusterRMP2005}, or even the adoption of
different algorithms ranging from quantum Monte Carlo to tensor
networks \cite{Shiwei2017,MEETS2021,Shiwei2023}. 
However, we emphasize that our results provide a clear
characterization of the entanglement properties of the
metallic and insulating solutions, which is expected to
be valid regardless of the approximation we used to identify it.

Different fruitful research directions can be envisaged in this
respect. Our information theory perspective can indeed provide precious information 
on the intriguing analogy between nonlocal correlations of the single-band Hubbard model and  correlations
between different local orbitals in multi-orbital systems~\cite{Capone2004PRL,Medici2014PRL,Werner2016PRB} and, in more general terms,
identify a conceptual framework to address the role of nonlocal correlations arising from local interactions in multi-component quantum systems including unconventional 
superconductors~\cite{Tomczak2012PRL}, correlated topological insulators~\cite{AA_BHZ_PRL,CrippaBHZ} and SU(N) cold-atom systems~\cite{Richaud2021PRB,Tusi2022NatPhys}.


\begin{acknowledgements}
The authors are thankful to C. Schilling and L. Ding for insightful discussions on
the formal definition of entanglement and correlation. 
A.A. acknowledges useful discussion with G. Sordi. 
G.B. further acknowledges illuminating comments from S. Giuli and M. Collura.
Funding is acknowledged by MUR through the PRIN 2017 (Prot.~20172H2SC4
005), PRIN 2020 (Prot.~2020JLZ52N 002) programs, the National Recovery and
Resilience Plan PNRR MUR (IT), PE0000023-NQSTI, financed by the European Union
- Next Generation EU and the MUR Italian National Centre for HPC, Big Data 
and Quantum Computing (grant number CN00000013) -  Mission 4 Component 2 Investments 1.3 and 1.4.
\end{acknowledgements}


\appendix

\section{Explicit formulas for superselected local entanglement, two-site total correlation and two-site entanglement}
\label{appendix:schillings_formulae}

In the following we will report the explicit expressions for the charge and parity 
\emph{superselected} measures of correlation and entanglement, as discussed in the
main body of the paper. Particularly we will address

\begin{itemize}
    \item the superselected entanglement between a single-site and its environment (Eqs.~\ref{eq:Ei_N-SSR} and \ref{eq:Ei_P-SSR}),
    \item the superselected total correlation between two neighboring sites (Eqs.~\ref{eq:Iij_N-SSR} and \ref{eq:Iij_P-SSR})
    \item the superselected entanglement between two neighboring sites (Eqs.~\ref{eq:Eij_N-SSR} and \ref{eq:Eij_P-SSR})
\end{itemize}

While fixing notations and language, we will also briefly comment on the physical 
meaning of these superselection rules and warn the reader about easy pitfalls on
the interpretation of the results, so to make our results as precise as we can.
Proofs will not be addressed but we will give explicit pointers to the relevant
literature.\\ 

Given the spin SU(2) and charge U(1) symmetries of the Hubbard model, the reduced 
density matrix for a single site $\rho_i$ is diagonal in the basis $\ket{\bullet}$,
$\ket{\uparrow}$, $\ket{\downarrow}$, $\ket{\uparrow\downarrow}$, where the black 
dot represents an empty site.
\begin{equation*}
    \rho_i = \begin{pmatrix}
        p_1 & 0 & 0 & 0\\
        0 & p_2 & 0 & 0\\
        0 & 0 & p_3 & 0\\
        0 & 0 & 0 & p_4\\
        \end{pmatrix}
\end{equation*}

If no superselection rule is taken into account and the ground state of the full lattice
is \emph{pure}, we can write it as Schmidt decomposition

\begin{align*}
    \ket{\Psi_\mathrm{gs}} = \enspace &\sqrt{p_1}\ket{\bullet}\enspace\otimes\enspace\ket{N,M} \,+ \\
     &\sqrt{p_2}\ket{\uparrow}\enspace\otimes\enspace\ket{N-1,M-\nicefrac{1}{2}} \,+ \\
     &\sqrt{p_3}\ket{\downarrow}\enspace\otimes\enspace\ket{N-1,M+\nicefrac{1}{2}} \,+ \\
     &\sqrt{p_4}\ket{\uparrow\downarrow}\otimes\enspace\ket{N-2,M},
\end{align*}

\noindent where $\ket{\mathcal{N},\mathcal{M}}$ represent a Fock state of $\mathcal{N}$
electrons and $\mathcal{M}$ magnetization, living on all lattice sites $j\neq i$. \\

In this case, the entanglement between the single site $i$ and the rest of
the lattice $\{j\neq i\}$, which we shall refer to as $E_i$, is just given by
the von Neumann entropy of $\rho_i$, namely
\begin{equation*}
    E_i \equiv s_i = -\sum_n p_n \log p_n.
\end{equation*}

However, as soon as we consider a superselection rule, namely a local restriction on the allowed
physical operators, such that all of them must commute with a given conserved quantity $\mathcal{Q}_i$,
the ground state $\Psi_\mathrm{gs}$ must be projected into the eigensubspaces of $\mathcal{Q}_i$, 
defining the block diagonal \emph{superselected} density matrix 
\begin{equation*}
    \rho_\mathrm{gs}^{\mathcal{Q}\text{-SSR}} = \sum_i\ketbra{\Psi_\mathrm{gs}^{\mathcal{Q}_i}}{\Psi_\mathrm{gs}^{\mathcal{Q}_i}}.
\end{equation*}

In general, the purity of $\rho_\mathrm{gs}^{\mathcal{Q}\text{-SSR}}$ cannot be guaranteed, so that all
its von Neumann reduced entropies (including the single-site entropy) are not legitimate measures of entanglement

\begin{equation*}
    E_i^{\mathcal{Q}\text{-SSR}} \neq -\sum_n p_n^{\mathcal{Q}\text{-SSR}} \log p_n^{\mathcal{Q}\text{-SSR}}.
\end{equation*}

A careful treatment, based on a general definition of entanglement as the minimal distance (quantum relative entropy) of
the given superselected local density matrix from the set of separable single-site states \cite{Vedral_RMP}, 
leads instead to the following expressions

\begin{align}
    E_i^\text{N-SSR} &= (p_2+p_3)\log(p_2+p_3) \label{eq:Ei_N-SSR}\\
    & \qquad -p_2\log p_2 - p_3\log p_3 \nonumber \\
    \nonumber \\
    E_i^\text{P-SSR} &= (p_1+p_4)\log(p_1+p_4) \label{eq:Ei_P-SSR}\\
    & \qquad + (p_2+p_3)\log(p_2+p_3) \nonumber\\
    & \qquad -p_1\log p_1 - p_2\log p_2 \nonumber\\
    & \qquad -p_3\log p_3 - p_4\log p_4 \nonumber
\end{align}

\noindent for the cases of $\mathcal{Q}_i = N_i$ and $\mathcal{Q}_i = P_i$,
being $N_i$ the local electron number and $P_i$ its parity. Notice that the
$p_n$ entering the two expressions are the original elements of the 
$\rho_i$ local density matrix, since it is already diagonal on the 
$N_i$ and $P_i$ sectors. The full derivation can be found in Ref.\,\cite{EntanglementVsCorrelation2021}.\\

\begin{figure}
    \includegraphics[width=0.475\textwidth]{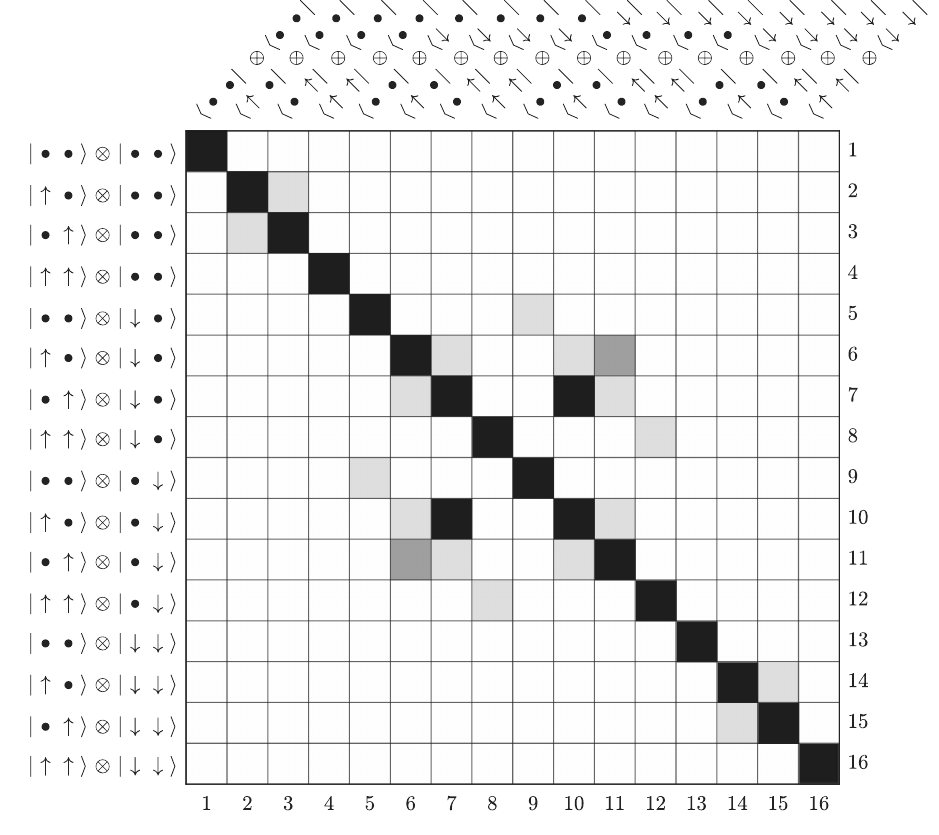}
    \caption{\label{fig:ssr_spy}
    Pictorial representation of the two-site reduced density matrix
    $\rho_{ij}$. For a generic interaction value $U/D$, most entries
    vanish (white) due to the spin and charge symmetries of the Hubbard
    model. Superselecting for local electron parity (P-SSR) sets all
    light-shaded entries to zero, while the N-SSR removes the remaining
    local charge fluctuations (the two dark-shaded entries). 
    Black elements are always preserved.}
\end{figure}

Let us move on a two-site subsystem, described by the $\rho_{ij}$ 
reduced density matrix. Fixing a basis for the two-site Fock space as
in Table~\ref{tab:dimer_basis}, we can define its matrix elements as
$p_{nm} = \mel{\psi_n}{\rho_{ij}}{\psi_m}$. Again, the spin SU(2) and 
charge U(1) symmetries of the Hubbard model impose several restrictions
on which $p_{nm}$ elements vanish, basically disallowing all fluctuations
in the two-site charge and magnetization.
Applying the number superselection rule (N-SSR) amounts to forbidding all 
changes in the local (single-site) charge, while the parity superselection
rule (P-SSR) allows only the charge fluctuations that preserve the parity of
the local particle number.
Local spin-flips survive both superselection rules, as they commute with the 
local particle number. A schematic depiction of the structure of $\rho_{ij}$,
and how it changes under N-SSR and P-SSR is reported in \figu{fig:ssr_spy}.
Once either the N-SSR or the P-SSR filtering is applied on $\rho_{ij}$,
the superselected nearest-neighbor total correlation is just given by 
the mutual information between its reduced single-site subsystems. In
functional terms, if we refer to the von Neumann entropy of a generic
density matrix $\rho$ as $s\{\rho\}$, we have
\begin{align}
    I_{ij}^\text{N-SSR} &= s\{\rho_i\}+s\{\rho_j\}-s\Bigl\{\rho_{ij}^\text{N-SSR}\Bigr\}, \label{eq:Iij_N-SSR}\\
    I_{ij}^\text{P-SSR} &= s\{\rho_i\}+s\{\rho_j\}-s\Bigl\{\rho_{ij}^\text{P-SSR}\Bigr\}. \label{eq:Iij_P-SSR}
\end{align}

We stress that the single-site reduced density matrices $\rho_i$, as
computed from $\rho_{ij}$ via the usual partial trace, are insensitive
to the N-SSR and P-SSR filtering, hence not superscripted in the formulas. 
Once again, note that while the superselected total correlation is still 
mathematically given by a subtraction of von Neumann entropies, the 
latter are not anymore legitimate measures of the entanglement between
single- / pairs of sites and their surroundings. Hence the superselected
total correlation cannot be physically interpreted as an algebraic sum
of superselected entanglement measures. An extended discussion on all these
subtle aspects can be found in Ref.\,\cite{Schilling_QuantumScience}.

\begin{table}
    \renewcommand{\arraystretch}{1.75} 
    \begin{tabular}{c||c}
        Label: $n$\enspace & \enspace State: $\ket{\psi_n^\uparrow} \otimes \ket{\psi_n^\downarrow}$ \\
        \hline
        1	& $\ket{\,\bullet\,\,\bullet\,}\otimes \ket{\,\bullet\,\,\bullet\,}$ \\
        2	& $\ket{ \,\uparrow\bullet\,}\otimes \ket{\,\bullet\,\,\bullet\,}$ \\
        3	& $\ket{\,\bullet\uparrow \,}\otimes \ket{\,\bullet\,\,\bullet\,}$ \\
        4	& $\ket{ \,\uparrow\,\,\uparrow\,}\otimes \ket{\,\bullet\,\,\bullet\,}$ \\
        5	& $\ket{\,\bullet\,\,\bullet\,}\otimes \ket{\,\downarrow\bullet\,}$ \\
        6	& $\ket{ \,\uparrow\bullet\,}\otimes \ket{\,\downarrow\bullet\,}$ \\
        7	& $\ket{\,\bullet\uparrow \,}\otimes \ket{\,\downarrow\bullet\,}$ \\
        8	& $\ket{ \,\uparrow\,\,\uparrow\,}\otimes \ket{\,\downarrow\bullet\,}$ \\
        9	& $\ket{\,\bullet\,\,\bullet\,}\otimes \ket{\,\bullet\downarrow \,}$ \\
        10	& $\ket{ \,\uparrow\bullet\,}\otimes \ket{\,\bullet\downarrow \,}$ \\
        11	& $\ket{\,\bullet\uparrow \,}\otimes \ket{\,\bullet\downarrow \,}$ \\
        12	& $\ket{ \,\uparrow\,\,\uparrow\,}\otimes \ket{\,\bullet\downarrow \,}$ \\
        13	& $\ket{\,\bullet\,\,\bullet\,}\otimes \ket{\,\downarrow\,\,\downarrow\,}$ \\
        14	& $\ket{ \,\uparrow\bullet\,}\otimes \ket{\,\downarrow\,\,\downarrow\,}$ \\
        15	& $\ket{\,\bullet\uparrow \,}\otimes \ket{\,\downarrow\,\,\downarrow\,}$ \\
        16	& $\ket{ \,\uparrow\,\,\uparrow\,}\otimes \ket{ \,\downarrow\,\,\downarrow\,}$ \\
    \end{tabular}
    \caption{\label{tab:dimer_basis}
    Basis for the two-site Fock space $\mathcal{F}_{ij}$. The indices $n$ of the quantum
    states $\ket{\psi_n} = \ket{\psi_n^\uparrow} \otimes \ket{\psi_n^\downarrow}$
    define the conventional labeling for the matrix elements $p_{nm}$ of the reduced
    density matrix for the two-site subsystem. 
    Black dots ($\bullet$) represent empty lattice sites.} 
\end{table}

\newpage

As discussed in the main body of the paper, we target the entanglement 
between a pair of local orbitals, defined as a suitable minimal ``distance'' 
(namely the quantum relative entropy) between the given reduced density matrix $\rho_{ij}$ and the set of separable
two-site states $\mathcal{D}_{ij}$ \cite{Vedral_RMP}. 
The relevant minimization process
constitutes a formidable open problem in quantum information theory, given that
no efficient description for the boundary of the $\mathcal{D}_{ij}$
set is known \cite{krueger2005open}. 
Recently a set of closed formulas has been
proposed, under the assumption of either N-SSR or P-SSR, plus some
typical symmetries of condensed matter systems \cite{Schilling_Entanglement_and_Locality_2023}.
In particular, we adopt here the versions valid under conservation of global $N$
and $S^z$, particle-hole symmetry and the assumption of a singlet ground state 
for the whole lattice, as they are indeed all satisfied by both the Hubbard model on the half-filled $2d$ square lattice and its CDMFT solution. \\
Here we report these two 
closed expressions, commenting on their physical properties with respect to the 
numerical results presented in \figu{fig:bond_entanglement}.
The reader interested in derivations and a general discussion on the relevance 
for arbitrary condensed matter systems and realistic quantum information processing 
therein, is strongly encouraged to approach the original references \cite{Schilling_Entanglement_and_Locality_2023,ding2023physical_entanglement}. \\


With reference to Table~\ref{tab:dimer_basis},
we define the following probabilities
\begin{align*}
    &p_n = \mel{\psi_n}{\rho_{ij}}{\psi_n}, \enspace\forall \,n\in[1,16]\setminus\{6,7,10,11\},\\[2mm]
    &\!\!\!\!\begin{pmatrix}
        p_6 \\
        p_{11}
    \end{pmatrix} = 
    R^{\dagger}
    \begin{pmatrix}
        \mel{\psi_6}{\rho_{ij}}{\psi_6} & \mel{\psi_{6}}{\rho_{ij}}{\psi_{11}} \\
        \mel{\psi_{11}}{\rho_{ij}}{\psi_{6}} & \mel{\psi_{11}}{\rho_{ij}}{\psi_{11}}
    \end{pmatrix}
    R,\\[2mm]
    &\!\!\!\!\begin{pmatrix}
        p_7 \\
        p_{10}
    \end{pmatrix} = 
    R^{\dagger}
    \begin{pmatrix}
        \mel{\psi_7}{\rho_{ij}}{\psi_7} & \mel{\psi_{7}}{\rho_{ij}}{\psi_{10}} \\
        \mel{\psi_{10}}{\rho_{ij}}{\psi_{7}} & \mel{\psi_{10}}{\rho_{ij}}{\psi_{10}}
    \end{pmatrix}
    R,\\[2mm]
    &\text{with}\enspace R =
    \begin{pmatrix}
        1 & -1 \\
        1 & +1
    \end{pmatrix} \times \frac{1}{\sqrt2}.
\end{align*}
Note that, according to symmetries, $\mel{\psi_7}{\rho_{ij}}{\psi_7}=\mel{\psi_{10}}{\rho_{ij}}{\psi_{10}}$
and $\mel{\psi_6}{\rho_{ij}}{\psi_6}=\mel{\psi_{11}}{\rho_{ij}}{\psi_{11}}$, 
so that the rotation matrix $R$ has been defined here as to 
diagonalize the subspaces defined by $\mathrm{Span}\{\ket{\psi_7},\ket{\psi_{10}}\}$
and $\mathrm{Span}\{\ket{\psi_6},\ket{\psi_{11}}\}$,
namely the only nondiagonal blocks surviving the superselection filtering on $\rho_{ij}$ 
(see \figu{fig:ssr_spy}). It can be further proven that the only relevant 
probabilities for determining the N-SSR and P-SSR two-site entanglement measures are
$p_1=p_{16}$ (equality ensured by particle-hole symmetry), 
$p_4=p_{13}$ (equality ensured by a singlet ground state), 
$p_6$, $p_{11}$, $p_7$ and $p_{10}$.\\

Finally we can express the charge and parity superselected two-site 
entanglement measures as 

\begin{align}
    E_{ij}^\text{N-SSR} &= \biggl[r \log\left(\frac{2r}{r+t}\right) + t\log\left(\frac{2t}{r+t}\right)\biggr] \times \theta(t-r), \label{eq:Eij_N-SSR}\\
    E_{ij}^\text{P-SSR} &= E_{ij}^\text{N-SSR} \enspace+ \nonumber \\
    &\enspace\,\biggl[ s \log\left(\frac{2s}{s+\tau}\right) + \tau\log\left(\frac{2\tau}{s+\tau}\right)\biggr] \times \theta(\tau-s), \label{eq:Eij_P-SSR}
\end{align}
\noindent where $\theta(t-r)$ and $\theta(\tau-s)$ are Heaviside steps, vanishing if $r \geq t$, and
$s \geq \tau$, respectively, where
\begin{align*}
    &t \,= \max\{p_7,p_{10}\}, \quad
    r    = \min\{p_7,p_{10}\} + p_4 + p_{13},\\
    &\tau = \max\{p_6,p_{11}\}, \quad
    s = \min\{p_6,p_{11}\} + p_1 + p_{16}.
\end{align*}
\noindent The vanishing imposed by the $\theta$-steps reflects the Peres--Horodecki
separability criterion \cite{Peres96,Horodecki97,HHHH_RMP}.\\

We highlight that Eqs.~\ref{eq:Eij_N-SSR} and \ref{eq:Eij_P-SSR} differ for a term depending on $\tau$ and $s$
alone, which involve only diagonal occupations of pairs of \emph{doublon} and \emph{holon} states. 
Since both of them are gradually suppressed by Mott localization we expect 
the two superselected measures of two-site entanglement to be asymptotically indistinguishable in 
the Mott insulator.\\


\section{Deriving a sum-rule for the two-site mutual information from the strong superadditivity property of quantum von Neumann entropies}
\label{appendix:subadditivity}

For any state of a generic quantum tripartite system 
$\mathcal{H}_\mathrm{ABC} = \mathcal{H}_\mathrm{A} \otimes \mathcal{H}_\mathrm{B} \otimes \mathcal{H}_\mathrm{C}$,
a \emph{strong subadditivity} property has been proven to relate the von Neumann entropies
of all its subsystems and the whole density matrix $\rho_\mathrm{ABC}$. 
In standard notation it reads \cite{Lieb_SSA}
\begin{equation}
s(\rho_\mathrm{ABC}) + s(\rho_\mathrm{B}) \leq s(\rho_\mathrm{AB}) + s(\rho_\mathrm{BC}).
\end{equation}

Adding the von Neumann entropy of subsystem A on both sides, we can recast the inequality 
in terms of suitable mutual informations, as
\begin{align}
    s(\rho_\mathrm{ABC}) + s(\rho_\mathrm{A}) + s(\rho_\mathrm{B}) &\leq s(\rho_\mathrm{AB}) + s(\rho_\mathrm{A}) + s(\rho_\mathrm{BC}) \nonumber \\
    \nonumber \\
    s(\rho_\mathrm{A}) + s(\rho_\mathrm{B}) - s(\rho_\mathrm{AB}) &\leq s(\rho_\mathrm{A}) + s(\rho_\mathrm{BC}) - s(\rho_\mathrm{ABC}) \nonumber \\
    \nonumber \\
    I(\rho_\mathrm{A}\vcentcolon\rho_\mathrm{B}) &\leq I(\rho_\mathrm{A}\vcentcolon\rho_\mathrm{BC})
    \label{eq:SSA_mutual_info}
\end{align}

Let us consider the CDMFT solution for the Hubbard model and take A and B as single-site orbitals $i$, $j$ in the cluster
and C as the rest of the impurity model. 
Equation \ref{eq:SSA_mutual_info} then becomes $I_{ij} \leq I_{i\{k\neq i\}}$.
Since the ground state of the impurity model is pure, we have $I_{i\{k\neq i\}} = s_i + s_{\{k\neq i\}} = 2s_i$
and we can further recast the inequality as 
\begin{equation}
    I_{ij} \leq 2s_i, \enspace\forall\,i\neq j
\end{equation}

Finally we fix a reference site $i$ and sum over all other possible cluster sites $j\neq i$,
to get
\begin{align}
    &\sum_{j=1}^{N_\mathrm{imp}} I_{ij}(1-\delta_{ij}) \leq \sum_{j=1}^{N_\mathrm{imp}} 2s_i(1-\delta_{ij}) \nonumber \\
    &\qquad \frac{\sum_{j=1}^{N_\mathrm{imp}} I_{ij}(1-\delta_{ij})}{2\sum_{j=1}^{N_\mathrm{imp}} (1-\delta_{ij})} \leq s_i
\end{align}
\noindent which is equivalent to Eq.~\ref{eq:worked_out_SSA} in the main text. \\

\section{Comparative table of entanglement markers for strongly correlated electrons}
\label{appendix:fordummies}

For ease of reference, here we provide a concise summary of all the correlation and entanglement markers
we discuss throughout this work.
We assume zero temperature and a general mixed state for the single- and two-site subsystems, due to the embedding in the lattice.

\begin{widetext}
\centering
\scalebox{0.95}
{
    \begin{tabular}{ll||c||c}
        \multicolumn{2}{c||}{\textsc{Marker}}   &\textsc{What it measures} & \textsc{What it bounds} \\
        \hline
        \Centerstack[l]{\\Single-site \\ entanglement \\ entropy (\ref{eq:local_entropy})\\} & 
        $\quad s_i$ &
        \Centerstack[c]{Entanglement between \\ a single site $i$ and \\ the rest of the lattice, \\ for pure ground states} & --- \\
        \hline
        \Centerstack[l]{\\Single-site \\entanglement \\ under charge \\ SSR (\ref{eq:Ei_N-SSR})\\} & 
        $\quad E^\text{N-SSR}_i$ &
        \Centerstack[c]{\\Accessible entanglement between \\ a single site $i$ and the rest of the \\ lattice, within operations that \\ conserve the local charge $N_i$\\} & 
        \Centerstack[c]{\\\enspace Entanglement between \enspace\\ a single site $i$ and the \\ rest of the lattice, \\ for all ground states \\ [from below]\\} \\
        \hline
        \Centerstack[l]{\\Single-site \\entanglement \\ under parity \\ SSR (\ref{eq:Ei_P-SSR})\\} & 
        $\quad E^\text{P-SSR}_i$ &
        \Centerstack[c]{\\\enspace Accessible entanglement between \enspace\\ a single site $i$ and the rest of the \\ lattice, within operations that \\ conserve the parity of the local \\ electron number $N_i$\\} & 
        \Centerstack[c]{Entanglement between \\ a single site $i$ and the \\ rest of the lattice, \\ for all ground states \\ [from below]} \\
        \hline
        \Centerstack[l]{\\Two-site \\ entanglement \\ entropy (\ref{eq:quasilocal_entropy})\\} &
        $\quad s_{ij}$&
        \Centerstack[c]{Entanglement between \\ the two sites $(i,j)$ and \\ the rest of the lattice, \\ for pure ground states} & --- \\
        \hline
        \Centerstack[l]{Two-site mutual \\ information (\ref{eq:bond_correlation})} &
        $\quad I_{ij}$ &
        \Centerstack[c]{Total correlation \\ between the two \\ sites $i$ and $j$} &
        \Centerstack[c]{\\The entanglement \\ between $i$ and $j$, \\ as well as any \\ two-site correlator \\ [from above]\\} \\
        \hline
        \Centerstack[l]{Two-site mutual \\ information  under \\ charge SSR (\ref{eq:Iij_N-SSR})} &
        $\quad I^\text{N-SSR}_{ij}$ &
        \Centerstack[c]{\\Accessible correlation \\ between the two \\ sites $i$ and $j$, within \\ operations that conserve \\ the local charge $N_i$\\} &
        \Centerstack[c]{The two-site mutual \\ information $I_{ij}$\\ [from below]} \\
        \hline
        \Centerstack[l]{Two-site mutual \\ information  under \\ parity SSR (\ref{eq:Iij_P-SSR})} &
        $\quad I^\text{P-SSR}_{ij}$ &
        \Centerstack[c]{\\Accessible correlation \\ between the two \\ sites $i$ and $j$, within \\ operations that conserve \\ the parity of the local \\ electron number $N_i$\\} &
        \Centerstack[c]{The two-site mutual \\ information $I_{ij}$\\ [from below]} \\
        \hline
        \Centerstack[l]{Two-site entanglement \\ under charge SSR (\ref{eq:Eij_N-SSR}) } &
        $\quad E^\text{N-SSR}_{ij}\quad$&
        \Centerstack[c]{\\Accessible entanglement \\ between $i$ and $j$, within \\ operations that conserve \\ the local charge $N_i$\\} & 
        \Centerstack[c]{The entanglement \\ between $i$ and $j$. \\ [from below]} \\
        \hline
        \Centerstack[l]{Two-site entanglement \\ under parity SSR (\ref{eq:Eij_P-SSR}) } &
        $\quad E^\text{P-SSR}_{ij}\quad$&
        \Centerstack[c]{\\Accessible entanglement \\ between $i$ and $j$, within \\ operations that conserve \\ the parity of the local \\ electron number $N_i$\\} & 
        \Centerstack[c]{The entanglement \\ between $i$ and $j$. \\ [from below]}
    \end{tabular}
}
\end{widetext}

\newpage

\section{Details of ASCI simulations}
\label{appendix:ASCI}

Here we provide some computational details of the ASCI calculations performed in this work for the different impurity models. 
As a reminder, a cluster with $N \times M$ correlated sites always has $6\times N\times M$ bath sites.

As discussed in Refs.\,\cite{Tubman2016,WilliamsYoung2023, ASCIdmft}, the important parameters for the ASCI solver include: (i) the size of the Hilbert space truncation, (ii) the size of the core space, namely the subset of determinants from which the Hilbert space exploration at each iteration is started, (iii) the active space threshold value, and (iv) the excitation threshold for the Green's function calculation, namely from which ground state determinants to build excited states in order to complete the determinant truncation to improve the Green's function accuracy.

The $1\times2$ cluster impurity model calculations were done without any Hilbert space truncation, while the $2\times2$ calculations were done with a Hilbert space truncation of $10^5$ determinants.
For the larger impurity models, namely $3\times2$ and $4\times2$, the simulations with $U/D > 0.25$ were also performed with $10^5$ determinants, while the $U/D = 0.25$ simulation was performed with $5\times10^5$ determinants to ensure convergence of Green's function and von Neumann entropies with respect to the truncation size.
In all cases, the core space from which to start the determinant search at each ASCI iteration (cf. Ref.\,\cite{ASCIdmft}) was set equal to the full truncation.
The threshold to define the active space in the approximate natural bath orbital basis was chosen to be $\delta = 10^{-4}$.
Once the ASCI ground state energy is converged, we evaluate the Green's function after complementing the determinant truncation as described in Ref.\,\cite{ASCIdmft,ASCIfit}, namely adding determinants corresponding to single electron excitations (within and beyond the active space) on top of those determinants in the ASCI truncation with absolute wave function coefficients larger than $10^{-4}$.
The DMFT self-consistency was iterated until the bath parameters were converged to an absolute error within $10^{-3}$, involving typically between 10 and 40 iterations.

\newpage


\bibliography{main_refs.bib,CMZ_refs.bib} 

\providecommand{\noopsort}[1]{}\providecommand{\singleletter}[1]{#1}%
\begin{thebibliography}{78}%
\makeatletter
\providecommand \@ifxundefined [1]{%
 \@ifx{#1\undefined}
}%
\providecommand \@ifnum [1]{%
 \ifnum #1\expandafter \@firstoftwo
 \else \expandafter \@secondoftwo
 \fi
}%
\providecommand \@ifx [1]{%
 \ifx #1\expandafter \@firstoftwo
 \else \expandafter \@secondoftwo
 \fi
}%
\providecommand \natexlab [1]{#1}%
\providecommand \enquote  [1]{``#1''}%
\providecommand \bibnamefont  [1]{#1}%
\providecommand \bibfnamefont [1]{#1}%
\providecommand \citenamefont [1]{#1}%
\providecommand \href@noop [0]{\@secondoftwo}%
\providecommand \href [0]{\begingroup \@sanitize@url \@href}%
\providecommand \@href[1]{\@@startlink{#1}\@@href}%
\providecommand \@@href[1]{\endgroup#1\@@endlink}%
\providecommand \@sanitize@url [0]{\catcode `\\12\catcode `\$12\catcode
  `\&12\catcode `\#12\catcode `\^12\catcode `\_12\catcode `\%12\relax}%
\providecommand \@@startlink[1]{}%
\providecommand \@@endlink[0]{}%
\providecommand \url  [0]{\begingroup\@sanitize@url \@url }%
\providecommand \@url [1]{\endgroup\@href {#1}{\urlprefix }}%
\providecommand \urlprefix  [0]{URL }%
\providecommand \Eprint [0]{\href }%
\providecommand \doibase [0]{https://doi.org/}%
\providecommand \selectlanguage [0]{\@gobble}%
\providecommand \bibinfo  [0]{\@secondoftwo}%
\providecommand \bibfield  [0]{\@secondoftwo}%
\providecommand \translation [1]{[#1]}%
\providecommand \BibitemOpen [0]{}%
\providecommand \bibitemStop [0]{}%
\providecommand \bibitemNoStop [0]{.\EOS\space}%
\providecommand \EOS [0]{\spacefactor3000\relax}%
\providecommand \BibitemShut  [1]{\csname bibitem#1\endcsname}%
\let\auto@bib@innerbib\@empty
\bibitem [{\citenamefont {{Mott}}(1968)}]{Mott_RMP}%
  \BibitemOpen
  \bibfield  {author} {\bibinfo {author} {\bibfnamefont {N.~F.}\ \bibnamefont
  {{Mott}}},\ }\bibfield  {title} {\bibinfo {title} {Metal-insulator
  transition},\ }\href {https://doi.org/10.1103/RevModPhys.40.677} {\bibfield
  {journal} {\bibinfo  {journal} {Rev. Mod. Phys.}\ }\textbf {\bibinfo {volume}
  {40}},\ \bibinfo {pages} {677} (\bibinfo {year} {1968})}\BibitemShut
  {NoStop}%
\bibitem [{\citenamefont {Imada}\ \emph {et~al.}(1998)\citenamefont {Imada},
  \citenamefont {Fujimori},\ and\ \citenamefont {Tokura}}]{MIT_RMP}%
  \BibitemOpen
  \bibfield  {author} {\bibinfo {author} {\bibfnamefont {M.}~\bibnamefont
  {Imada}}, \bibinfo {author} {\bibfnamefont {A.}~\bibnamefont {Fujimori}},\
  and\ \bibinfo {author} {\bibfnamefont {Y.}~\bibnamefont {Tokura}},\
  }\bibfield  {title} {\bibinfo {title} {Metal-insulator transitions},\ }\href
  {https://doi.org/10.1103/RevModPhys.70.1039} {\bibfield  {journal} {\bibinfo
  {journal} {Rev. Mod. Phys.}\ }\textbf {\bibinfo {volume} {70}},\ \bibinfo
  {pages} {1039} (\bibinfo {year} {1998})}\BibitemShut {NoStop}%
\bibitem [{\citenamefont {Osterloh}\ \emph {et~al.}(2002)\citenamefont
  {Osterloh}, \citenamefont {Amico}, \citenamefont {Falci},\ and\ \citenamefont
  {Fazio}}]{Osterloh_Nature}%
  \BibitemOpen
  \bibfield  {author} {\bibinfo {author} {\bibfnamefont {A.}~\bibnamefont
  {Osterloh}}, \bibinfo {author} {\bibfnamefont {L.}~\bibnamefont {Amico}},
  \bibinfo {author} {\bibfnamefont {G.}~\bibnamefont {Falci}},\ and\ \bibinfo
  {author} {\bibfnamefont {R.}~\bibnamefont {Fazio}},\ }\bibfield  {title}
  {\bibinfo {title} {Scaling of entanglement close to a quantum phase
  transition},\ }\href {https://doi.org/10.1038/416608a} {\bibfield  {journal}
  {\bibinfo  {journal} {Nature}\ }\textbf {\bibinfo {volume} {416}},\ \bibinfo
  {pages} {608} (\bibinfo {year} {2002})}\BibitemShut {NoStop}%
\bibitem [{\citenamefont {Osborne}\ and\ \citenamefont
  {Nielsen}(2002)}]{QuasiocalEntanglement_simpleQPT}%
  \BibitemOpen
  \bibfield  {author} {\bibinfo {author} {\bibfnamefont {T.~J.}\ \bibnamefont
  {Osborne}}\ and\ \bibinfo {author} {\bibfnamefont {M.~A.}\ \bibnamefont
  {Nielsen}},\ }\bibfield  {title} {\bibinfo {title} {Entanglement in a simple
  quantum phase transition},\ }\href
  {https://doi.org/10.1103/PhysRevA.66.032110} {\bibfield  {journal} {\bibinfo
  {journal} {Phys. Rev. A}\ }\textbf {\bibinfo {volume} {66}},\ \bibinfo
  {pages} {032110} (\bibinfo {year} {2002})}\BibitemShut {NoStop}%
\bibitem [{\citenamefont {Vidal}\ \emph {et~al.}(2003)\citenamefont {Vidal},
  \citenamefont {Latorre}, \citenamefont {Rico},\ and\ \citenamefont
  {Kitaev}}]{Vidal_Entanglement_in_QCP}%
  \BibitemOpen
  \bibfield  {author} {\bibinfo {author} {\bibfnamefont {G.}~\bibnamefont
  {Vidal}}, \bibinfo {author} {\bibfnamefont {J.~I.}\ \bibnamefont {Latorre}},
  \bibinfo {author} {\bibfnamefont {E.}~\bibnamefont {Rico}},\ and\ \bibinfo
  {author} {\bibfnamefont {A.}~\bibnamefont {Kitaev}},\ }\bibfield  {title}
  {\bibinfo {title} {Entanglement in quantum critical phenomena},\ }\href
  {https://doi.org/10.1103/PhysRevLett.90.227902} {\bibfield  {journal}
  {\bibinfo  {journal} {Phys. Rev. Lett.}\ }\textbf {\bibinfo {volume} {90}},\
  \bibinfo {pages} {227902} (\bibinfo {year} {2003})}\BibitemShut {NoStop}%
\bibitem [{\citenamefont {Squillante}\ \emph {et~al.}(2023)\citenamefont
  {Squillante}, \citenamefont {Ricco}, \citenamefont {Ukpong}, \citenamefont
  {Lagos-Monaco}, \citenamefont {Seridonio},\ and\ \citenamefont
  {de~Souza}}]{squillante2023gruneisen}%
  \BibitemOpen
  \bibfield  {author} {\bibinfo {author} {\bibfnamefont {L.}~\bibnamefont
  {Squillante}}, \bibinfo {author} {\bibfnamefont {L.~S.}\ \bibnamefont
  {Ricco}}, \bibinfo {author} {\bibfnamefont {A.~M.}\ \bibnamefont {Ukpong}},
  \bibinfo {author} {\bibfnamefont {R.~E.}\ \bibnamefont {Lagos-Monaco}},
  \bibinfo {author} {\bibfnamefont {A.~C.}\ \bibnamefont {Seridonio}},\ and\
  \bibinfo {author} {\bibfnamefont {M.}~\bibnamefont {de~Souza}},\ }\bibfield
  {title} {\bibinfo {title} {Gr\"uneisen parameter as an entanglement compass
  and the breakdown of the hellmann-feynman theorem},\ }\href
  {https://doi.org/10.1103/PhysRevB.108.L140403} {\bibfield  {journal}
  {\bibinfo  {journal} {Phys. Rev. B}\ }\textbf {\bibinfo {volume} {108}},\
  \bibinfo {pages} {L140403} (\bibinfo {year} {2023})}\BibitemShut {NoStop}%
\bibitem [{\citenamefont {Kitaev}\ and\ \citenamefont
  {Preskill}(2006)}]{Kitaev_TopologicalEntropy}%
  \BibitemOpen
  \bibfield  {author} {\bibinfo {author} {\bibfnamefont {A.}~\bibnamefont
  {Kitaev}}\ and\ \bibinfo {author} {\bibfnamefont {J.}~\bibnamefont
  {Preskill}},\ }\bibfield  {title} {\bibinfo {title} {Topological entanglement
  entropy},\ }\href {https://doi.org/10.1103/PhysRevLett.96.110404} {\bibfield
  {journal} {\bibinfo  {journal} {Phys. Rev. Lett.}\ }\textbf {\bibinfo
  {volume} {96}},\ \bibinfo {pages} {110404} (\bibinfo {year}
  {2006})}\BibitemShut {NoStop}%
\bibitem [{\citenamefont {Levin}\ and\ \citenamefont
  {Wen}(2006)}]{Levin_TopologicalOrder}%
  \BibitemOpen
  \bibfield  {author} {\bibinfo {author} {\bibfnamefont {M.}~\bibnamefont
  {Levin}}\ and\ \bibinfo {author} {\bibfnamefont {X.-G.}\ \bibnamefont
  {Wen}},\ }\bibfield  {title} {\bibinfo {title} {Detecting topological order
  in a ground state wave function},\ }\href
  {https://doi.org/10.1103/PhysRevLett.96.110405} {\bibfield  {journal}
  {\bibinfo  {journal} {Phys. Rev. Lett.}\ }\textbf {\bibinfo {volume} {96}},\
  \bibinfo {pages} {110405} (\bibinfo {year} {2006})}\BibitemShut {NoStop}%
\bibitem [{\citenamefont {Chen}\ \emph {et~al.}(2010)\citenamefont {Chen},
  \citenamefont {Gu},\ and\ \citenamefont {Wen}}]{Wen_TopologicalOrder}%
  \BibitemOpen
  \bibfield  {author} {\bibinfo {author} {\bibfnamefont {X.}~\bibnamefont
  {Chen}}, \bibinfo {author} {\bibfnamefont {Z.-C.}\ \bibnamefont {Gu}},\ and\
  \bibinfo {author} {\bibfnamefont {X.-G.}\ \bibnamefont {Wen}},\ }\bibfield
  {title} {\bibinfo {title} {Local unitary transformation, long-range quantum
  entanglement, wave function renormalization, and topological order},\ }\href
  {https://doi.org/10.1103/PhysRevB.82.155138} {\bibfield  {journal} {\bibinfo
  {journal} {Phys. Rev. B}\ }\textbf {\bibinfo {volume} {82}},\ \bibinfo
  {pages} {155138} (\bibinfo {year} {2010})}\BibitemShut {NoStop}%
\bibitem [{\citenamefont {Boguslawski}\ \emph {et~al.}(2013)\citenamefont
  {Boguslawski}, \citenamefont {Tecmer}, \citenamefont {Barcza}, \citenamefont
  {Legeza},\ and\ \citenamefont {Reiher}}]{Entanglement_and_Bonding}%
  \BibitemOpen
  \bibfield  {author} {\bibinfo {author} {\bibfnamefont {K.}~\bibnamefont
  {Boguslawski}}, \bibinfo {author} {\bibfnamefont {P.}~\bibnamefont {Tecmer}},
  \bibinfo {author} {\bibfnamefont {G.}~\bibnamefont {Barcza}}, \bibinfo
  {author} {\bibfnamefont {O.}~\bibnamefont {Legeza}},\ and\ \bibinfo {author}
  {\bibfnamefont {M.}~\bibnamefont {Reiher}},\ }\bibfield  {title} {\bibinfo
  {title} {Orbital entanglement in bond-formation processes},\ }\href
  {https://doi.org/10.1021/ct400247p} {\bibfield  {journal} {\bibinfo
  {journal} {Journal of Chemical Theory and Computation}\ }\textbf {\bibinfo
  {volume} {9}},\ \bibinfo {pages} {2959} (\bibinfo {year} {2013})},\ \bibinfo
  {note} {pMID: 26583979}\BibitemShut {NoStop}%
\bibitem [{\citenamefont {Szalay}\ \emph {et~al.}(2017)\citenamefont {Szalay},
  \citenamefont {Barcza}, \citenamefont {Szilvasi}, \citenamefont {Veis},\ and\
  \citenamefont {Legeza}}]{Correlation_and_bonding}%
  \BibitemOpen
  \bibfield  {author} {\bibinfo {author} {\bibfnamefont {S.}~\bibnamefont
  {Szalay}}, \bibinfo {author} {\bibfnamefont {G.}~\bibnamefont {Barcza}},
  \bibinfo {author} {\bibfnamefont {T.}~\bibnamefont {Szilvasi}}, \bibinfo
  {author} {\bibfnamefont {L.}~\bibnamefont {Veis}},\ and\ \bibinfo {author}
  {\bibfnamefont {O.}~\bibnamefont {Legeza}},\ }\bibfield  {title} {\bibinfo
  {title} {The correlation theory of the chemical bond},\ }\href
  {https://doi.org/10.1038/s41598-017-02447-z} {\bibfield  {journal} {\bibinfo
  {journal} {Scientific Reports}\ }\textbf {\bibinfo {volume} {7}},\ \bibinfo
  {pages} {2237} (\bibinfo {year} {2017})}\BibitemShut {NoStop}%
\bibitem [{\citenamefont {Ding}\ \emph
  {et~al.}(2022{\natexlab{a}})\citenamefont {Ding}, \citenamefont {Knecht},
  \citenamefont {Zimborás},\ and\ \citenamefont
  {Schilling}}]{Schilling_QuantumScience}%
  \BibitemOpen
  \bibfield  {author} {\bibinfo {author} {\bibfnamefont {L.}~\bibnamefont
  {Ding}}, \bibinfo {author} {\bibfnamefont {S.}~\bibnamefont {Knecht}},
  \bibinfo {author} {\bibfnamefont {Z.}~\bibnamefont {Zimborás}},\ and\
  \bibinfo {author} {\bibfnamefont {C.}~\bibnamefont {Schilling}},\ }\bibfield
  {title} {\bibinfo {title} {Quantum correlations in molecules: from quantum
  resourcing to chemical bonding},\ }\href
  {https://doi.org/10.1088/2058-9565/aca4ee} {\bibfield  {journal} {\bibinfo
  {journal} {Quantum Science and Technology}\ }\textbf {\bibinfo {volume}
  {8}},\ \bibinfo {pages} {015015} (\bibinfo {year}
  {2022}{\natexlab{a}})}\BibitemShut {NoStop}%
\bibitem [{\citenamefont {White}(1992)}]{White_DMRG}%
  \BibitemOpen
  \bibfield  {author} {\bibinfo {author} {\bibfnamefont {S.~R.}\ \bibnamefont
  {White}},\ }\bibfield  {title} {\bibinfo {title} {Density matrix formulation
  for quantum renormalization groups},\ }\href
  {https://doi.org/10.1103/PhysRevLett.69.2863} {\bibfield  {journal} {\bibinfo
   {journal} {Phys. Rev. Lett.}\ }\textbf {\bibinfo {volume} {69}},\ \bibinfo
  {pages} {2863} (\bibinfo {year} {1992})}\BibitemShut {NoStop}%
\bibitem [{\citenamefont {Legeza}\ and\ \citenamefont
  {S\'olyom}(2003)}]{Optimizing_DMRG}%
  \BibitemOpen
  \bibfield  {author} {\bibinfo {author} {\bibfnamefont {O.}~\bibnamefont
  {Legeza}}\ and\ \bibinfo {author} {\bibfnamefont {J.}~\bibnamefont
  {S\'olyom}},\ }\bibfield  {title} {\bibinfo {title} {Optimizing the
  density-matrix renormalization group method using quantum information
  entropy},\ }\href {https://doi.org/10.1103/PhysRevB.68.195116} {\bibfield
  {journal} {\bibinfo  {journal} {Phys. Rev. B}\ }\textbf {\bibinfo {volume}
  {68}},\ \bibinfo {pages} {195116} (\bibinfo {year} {2003})}\BibitemShut
  {NoStop}%
\bibitem [{\citenamefont {Stein}\ and\ \citenamefont
  {Reiher}(2016)}]{Automated_Active_Spaces}%
  \BibitemOpen
  \bibfield  {author} {\bibinfo {author} {\bibfnamefont {C.~J.}\ \bibnamefont
  {Stein}}\ and\ \bibinfo {author} {\bibfnamefont {M.}~\bibnamefont {Reiher}},\
  }\bibfield  {title} {\bibinfo {title} {Automated selection of active orbital
  spaces},\ }\href {https://doi.org/10.1021/acs.jctc.6b00156} {\bibfield
  {journal} {\bibinfo  {journal} {Journal of Chemical Theory and Computation}\
  }\textbf {\bibinfo {volume} {12}},\ \bibinfo {pages} {1760} (\bibinfo {year}
  {2016})}\BibitemShut {NoStop}%
\bibitem [{\citenamefont {Georges}\ \emph {et~al.}(1996)\citenamefont
  {Georges}, \citenamefont {Kotliar}, \citenamefont {Krauth},\ and\
  \citenamefont {Rozenberg}}]{DMFT_RMP}%
  \BibitemOpen
  \bibfield  {author} {\bibinfo {author} {\bibfnamefont {A.}~\bibnamefont
  {Georges}}, \bibinfo {author} {\bibfnamefont {G.}~\bibnamefont {Kotliar}},
  \bibinfo {author} {\bibfnamefont {W.}~\bibnamefont {Krauth}},\ and\ \bibinfo
  {author} {\bibfnamefont {M.~J.}\ \bibnamefont {Rozenberg}},\ }\bibfield
  {title} {\bibinfo {title} {Dynamical mean-field theory of strongly correlated
  fermion systems and the limit of infinite dimensions},\ }\href
  {https://doi.org/10.1103/RevModPhys.68.13} {\bibfield  {journal} {\bibinfo
  {journal} {Rev. Mod. Phys.}\ }\textbf {\bibinfo {volume} {68}},\ \bibinfo
  {pages} {13} (\bibinfo {year} {1996})}\BibitemShut {NoStop}%
\bibitem [{\citenamefont {Su}\ \emph {et~al.}(2013)\citenamefont {Su},
  \citenamefont {Dai},\ and\ \citenamefont {Tong}}]{XiDai2013}%
  \BibitemOpen
  \bibfield  {author} {\bibinfo {author} {\bibfnamefont {D.~D.}\ \bibnamefont
  {Su}}, \bibinfo {author} {\bibfnamefont {X.}~\bibnamefont {Dai}},\ and\
  \bibinfo {author} {\bibfnamefont {N.~H.}\ \bibnamefont {Tong}},\ }\bibfield
  {title} {\bibinfo {title} {Local entanglement entropy at the {Mott}
  metal-insulator transition in infinite dimensions},\ }\href
  {https://doi.org/10.1142/S0217984913500346} {\bibfield  {journal} {\bibinfo
  {journal} {Modern Physics Letters B}\ }\textbf {\bibinfo {volume} {27}},\
  \bibinfo {pages} {1350034} (\bibinfo {year} {2013})}\BibitemShut {NoStop}%
\bibitem [{\citenamefont {Walsh}\ \emph {et~al.}(2019)\citenamefont {Walsh},
  \citenamefont {S\'emon}, \citenamefont {Poulin}, \citenamefont {Sordi},\ and\
  \citenamefont {Tremblay}}]{WalshLetter2019}%
  \BibitemOpen
  \bibfield  {author} {\bibinfo {author} {\bibfnamefont {C.}~\bibnamefont
  {Walsh}}, \bibinfo {author} {\bibfnamefont {P.}~\bibnamefont {S\'emon}},
  \bibinfo {author} {\bibfnamefont {D.}~\bibnamefont {Poulin}}, \bibinfo
  {author} {\bibfnamefont {G.}~\bibnamefont {Sordi}},\ and\ \bibinfo {author}
  {\bibfnamefont {A.-M.~S.}\ \bibnamefont {Tremblay}},\ }\bibfield  {title}
  {\bibinfo {title} {Local entanglement entropy and mutual information across
  the {Mott} transition in the two-dimensional {Hubbard} model},\ }\href
  {https://doi.org/10.1103/PhysRevLett.122.067203} {\bibfield  {journal}
  {\bibinfo  {journal} {Phys. Rev. Lett.}\ }\textbf {\bibinfo {volume} {122}},\
  \bibinfo {pages} {067203} (\bibinfo {year} {2019})}\BibitemShut {NoStop}%
\bibitem [{\citenamefont {Walsh}\ \emph {et~al.}(2020)\citenamefont {Walsh},
  \citenamefont {S\'emon}, \citenamefont {Poulin}, \citenamefont {Sordi},\ and\
  \citenamefont {Tremblay}}]{WalshPRXQuantum2020}%
  \BibitemOpen
  \bibfield  {author} {\bibinfo {author} {\bibfnamefont {C.}~\bibnamefont
  {Walsh}}, \bibinfo {author} {\bibfnamefont {P.}~\bibnamefont {S\'emon}},
  \bibinfo {author} {\bibfnamefont {D.}~\bibnamefont {Poulin}}, \bibinfo
  {author} {\bibfnamefont {G.}~\bibnamefont {Sordi}},\ and\ \bibinfo {author}
  {\bibfnamefont {A.-M.~S.}\ \bibnamefont {Tremblay}},\ }\bibfield  {title}
  {\bibinfo {title} {Entanglement and classical correlations at the
  doping-driven {Mott} transition in the two-dimensional {Hubbard} model},\
  }\href {https://doi.org/10.1103/PRXQuantum.1.020310} {\bibfield  {journal}
  {\bibinfo  {journal} {PRX Quantum}\ }\textbf {\bibinfo {volume} {1}},\
  \bibinfo {pages} {020310} (\bibinfo {year} {2020})}\BibitemShut {NoStop}%
\bibitem [{\citenamefont {Walsh}\ \emph {et~al.}(2021)\citenamefont {Walsh},
  \citenamefont {Charlebois}, \citenamefont {S{\'{e}}mon}, \citenamefont
  {Sordi},\ and\ \citenamefont {Tremblay}}]{WalshPNAS2021}%
  \BibitemOpen
  \bibfield  {author} {\bibinfo {author} {\bibfnamefont {C.}~\bibnamefont
  {Walsh}}, \bibinfo {author} {\bibfnamefont {M.}~\bibnamefont {Charlebois}},
  \bibinfo {author} {\bibfnamefont {P.}~\bibnamefont {S{\'{e}}mon}}, \bibinfo
  {author} {\bibfnamefont {G.}~\bibnamefont {Sordi}},\ and\ \bibinfo {author}
  {\bibfnamefont {A.-M.~S.}\ \bibnamefont {Tremblay}},\ }\bibfield  {title}
  {\bibinfo {title} {Information-theoretic measures of superconductivity in a
  two-dimensional doped {Mott} insulator},\ }\href
  {https://doi.org/10.1073/pnas.2104114118} {\bibfield  {journal} {\bibinfo
  {journal} {Proceedings of the National Academy of Sciences}\ }\textbf
  {\bibinfo {volume} {118}},\ \bibinfo {pages} {e2104114118} (\bibinfo {year}
  {2021})}\BibitemShut {NoStop}%
\bibitem [{\citenamefont {Kotliar}\ \emph {et~al.}(2001)\citenamefont
  {Kotliar}, \citenamefont {Savrasov}, \citenamefont {P\'alsson},\ and\
  \citenamefont {Biroli}}]{Kotliar2001}%
  \BibitemOpen
  \bibfield  {author} {\bibinfo {author} {\bibfnamefont {G.}~\bibnamefont
  {Kotliar}}, \bibinfo {author} {\bibfnamefont {S.~Y.}\ \bibnamefont
  {Savrasov}}, \bibinfo {author} {\bibfnamefont {G.}~\bibnamefont
  {P\'alsson}},\ and\ \bibinfo {author} {\bibfnamefont {G.}~\bibnamefont
  {Biroli}},\ }\bibfield  {title} {\bibinfo {title} {Cellular dynamical mean
  field approach to strongly correlated systems},\ }\href
  {https://doi.org/10.1103/PhysRevLett.87.186401} {\bibfield  {journal}
  {\bibinfo  {journal} {Phys. Rev. Lett.}\ }\textbf {\bibinfo {volume} {87}},\
  \bibinfo {pages} {186401} (\bibinfo {year} {2001})}\BibitemShut {NoStop}%
\bibitem [{\citenamefont {Maier}\ \emph {et~al.}(2005)\citenamefont {Maier},
  \citenamefont {Jarrell}, \citenamefont {Pruschke},\ and\ \citenamefont
  {Hettler}}]{ClusterRMP2005}%
  \BibitemOpen
  \bibfield  {author} {\bibinfo {author} {\bibfnamefont {T.}~\bibnamefont
  {Maier}}, \bibinfo {author} {\bibfnamefont {M.}~\bibnamefont {Jarrell}},
  \bibinfo {author} {\bibfnamefont {T.}~\bibnamefont {Pruschke}},\ and\
  \bibinfo {author} {\bibfnamefont {M.~H.}\ \bibnamefont {Hettler}},\
  }\bibfield  {title} {\bibinfo {title} {Quantum cluster theories},\ }\href
  {https://doi.org/10.1103/RevModPhys.77.1027} {\bibfield  {journal} {\bibinfo
  {journal} {Rev. Mod. Phys.}\ }\textbf {\bibinfo {volume} {77}},\ \bibinfo
  {pages} {1027} (\bibinfo {year} {2005})}\BibitemShut {NoStop}%
\bibitem [{\citenamefont {Park}\ \emph {et~al.}(2008)\citenamefont {Park},
  \citenamefont {Haule},\ and\ \citenamefont {Kotliar}}]{Haule2008}%
  \BibitemOpen
  \bibfield  {author} {\bibinfo {author} {\bibfnamefont {H.}~\bibnamefont
  {Park}}, \bibinfo {author} {\bibfnamefont {K.}~\bibnamefont {Haule}},\ and\
  \bibinfo {author} {\bibfnamefont {G.}~\bibnamefont {Kotliar}},\ }\bibfield
  {title} {\bibinfo {title} {Cluster dynamical mean field theory of the {Mott}
  transition},\ }\href {https://doi.org/10.1103/PhysRevLett.101.186403}
  {\bibfield  {journal} {\bibinfo  {journal} {Phys. Rev. Lett.}\ }\textbf
  {\bibinfo {volume} {101}},\ \bibinfo {pages} {186403} (\bibinfo {year}
  {2008})}\BibitemShut {NoStop}%
\bibitem [{\citenamefont {Cocchi}\ \emph {et~al.}(2017)\citenamefont {Cocchi},
  \citenamefont {Miller}, \citenamefont {Drewes}, \citenamefont {Chan},
  \citenamefont {Pertot}, \citenamefont {Brennecke},\ and\ \citenamefont
  {K\"ohl}}]{Cocchi_ColdAtoms}%
  \BibitemOpen
  \bibfield  {author} {\bibinfo {author} {\bibfnamefont {E.}~\bibnamefont
  {Cocchi}}, \bibinfo {author} {\bibfnamefont {L.~A.}\ \bibnamefont {Miller}},
  \bibinfo {author} {\bibfnamefont {J.~H.}\ \bibnamefont {Drewes}}, \bibinfo
  {author} {\bibfnamefont {C.~F.}\ \bibnamefont {Chan}}, \bibinfo {author}
  {\bibfnamefont {D.}~\bibnamefont {Pertot}}, \bibinfo {author} {\bibfnamefont
  {F.}~\bibnamefont {Brennecke}},\ and\ \bibinfo {author} {\bibfnamefont
  {M.}~\bibnamefont {K\"ohl}},\ }\bibfield  {title} {\bibinfo {title}
  {Measuring entropy and short-range correlations in the two-dimensional
  {Hubbard} model},\ }\href {https://doi.org/10.1103/PhysRevX.7.031025}
  {\bibfield  {journal} {\bibinfo  {journal} {Phys. Rev. X}\ }\textbf {\bibinfo
  {volume} {7}},\ \bibinfo {pages} {031025} (\bibinfo {year}
  {2017})}\BibitemShut {NoStop}%
\bibitem [{\citenamefont {Amico}\ \emph {et~al.}(2008)\citenamefont {Amico},
  \citenamefont {Fazio}, \citenamefont {Osterloh},\ and\ \citenamefont
  {Vedral}}]{EntanglementRMP}%
  \BibitemOpen
  \bibfield  {author} {\bibinfo {author} {\bibfnamefont {L.}~\bibnamefont
  {Amico}}, \bibinfo {author} {\bibfnamefont {R.}~\bibnamefont {Fazio}},
  \bibinfo {author} {\bibfnamefont {A.}~\bibnamefont {Osterloh}},\ and\
  \bibinfo {author} {\bibfnamefont {V.}~\bibnamefont {Vedral}},\ }\bibfield
  {title} {\bibinfo {title} {Entanglement in many-body systems},\ }\href
  {https://doi.org/10.1103/RevModPhys.80.517} {\bibfield  {journal} {\bibinfo
  {journal} {Rev. Mod. Phys.}\ }\textbf {\bibinfo {volume} {80}},\ \bibinfo
  {pages} {517} (\bibinfo {year} {2008})}\BibitemShut {NoStop}%
\bibitem [{\citenamefont {Bera}\ \emph {et~al.}(2024)\citenamefont {Bera},
  \citenamefont {Haldar},\ and\ \citenamefont {Banerjee}}]{Renyi_DMFT}%
  \BibitemOpen
  \bibfield  {author} {\bibinfo {author} {\bibfnamefont {S.}~\bibnamefont
  {Bera}}, \bibinfo {author} {\bibfnamefont {A.}~\bibnamefont {Haldar}},\ and\
  \bibinfo {author} {\bibfnamefont {S.}~\bibnamefont {Banerjee}},\ }\bibfield
  {title} {\bibinfo {title} {Dynamical mean-field theory for {R\'enyi}
  entanglement entropy and mutual information in the hubbard model},\ }\href
  {https://doi.org/10.1103/PhysRevB.109.035156} {\bibfield  {journal} {\bibinfo
   {journal} {Phys. Rev. B}\ }\textbf {\bibinfo {volume} {109}},\ \bibinfo
  {pages} {035156} (\bibinfo {year} {2024})}\BibitemShut {NoStop}%
\bibitem [{\citenamefont {D'Emidio}\ \emph {et~al.}(2024)\citenamefont
  {D'Emidio}, \citenamefont {Or\'us}, \citenamefont {Laflorencie},\ and\
  \citenamefont {de~Juan}}]{Renyi_Honeycomb}%
  \BibitemOpen
  \bibfield  {author} {\bibinfo {author} {\bibfnamefont {J.}~\bibnamefont
  {D'Emidio}}, \bibinfo {author} {\bibfnamefont {R.}~\bibnamefont {Or\'us}},
  \bibinfo {author} {\bibfnamefont {N.}~\bibnamefont {Laflorencie}},\ and\
  \bibinfo {author} {\bibfnamefont {F.}~\bibnamefont {de~Juan}},\ }\bibfield
  {title} {\bibinfo {title} {Universal features of entanglement entropy in the
  honeycomb hubbard model},\ }\href
  {https://doi.org/10.1103/PhysRevLett.132.076502} {\bibfield  {journal}
  {\bibinfo  {journal} {Phys. Rev. Lett.}\ }\textbf {\bibinfo {volume} {132}},\
  \bibinfo {pages} {076502} (\bibinfo {year} {2024})}\BibitemShut {NoStop}%
\bibitem [{\citenamefont {Capone}\ \emph {et~al.}(2007)\citenamefont {Capone},
  \citenamefont {de' Medici},\ and\ \citenamefont {Georges}}]{DMFT/ED_Capone}%
  \BibitemOpen
  \bibfield  {author} {\bibinfo {author} {\bibfnamefont {M.}~\bibnamefont
  {Capone}}, \bibinfo {author} {\bibfnamefont {L.}~\bibnamefont {de' Medici}},\
  and\ \bibinfo {author} {\bibfnamefont {A.}~\bibnamefont {Georges}},\
  }\bibfield  {title} {\bibinfo {title} {Solving the dynamical mean-field
  theory at very low temperatures using the {Lanczos} exact diagonalization},\
  }\href {https://doi.org/10.1103/PhysRevB.76.245116} {\bibfield  {journal}
  {\bibinfo  {journal} {Phys. Rev. B}\ }\textbf {\bibinfo {volume} {76}},\
  \bibinfo {pages} {245116} (\bibinfo {year} {2007})}\BibitemShut {NoStop}%
\bibitem [{\citenamefont {Amaricci}\ \emph {et~al.}(2022)\citenamefont
  {Amaricci}, \citenamefont {Crippa}, \citenamefont {Scazzola}, \citenamefont
  {Petocchi}, \citenamefont {Mazza}, \citenamefont {{de Medici}},\ and\
  \citenamefont {Capone}}]{EDIpack}%
  \BibitemOpen
  \bibfield  {author} {\bibinfo {author} {\bibfnamefont {A.}~\bibnamefont
  {Amaricci}}, \bibinfo {author} {\bibfnamefont {L.}~\bibnamefont {Crippa}},
  \bibinfo {author} {\bibfnamefont {A.}~\bibnamefont {Scazzola}}, \bibinfo
  {author} {\bibfnamefont {F.}~\bibnamefont {Petocchi}}, \bibinfo {author}
  {\bibfnamefont {G.}~\bibnamefont {Mazza}}, \bibinfo {author} {\bibfnamefont
  {L.}~\bibnamefont {{de Medici}}},\ and\ \bibinfo {author} {\bibfnamefont
  {M.}~\bibnamefont {Capone}},\ }\bibfield  {title} {\bibinfo {title}
  {{EDIpack}: A parallel exact diagonalization package for quantum impurity
  problems},\ }\href
  {https://doi.org/https://doi.org/10.1016/j.cpc.2021.108261} {\bibfield
  {journal} {\bibinfo  {journal} {Computer Physics Communications}\ }\textbf
  {\bibinfo {volume} {273}},\ \bibinfo {pages} {108261} (\bibinfo {year}
  {2022})}\BibitemShut {NoStop}%
\bibitem [{\citenamefont {Mejuto-Zaera}\ \emph {et~al.}(2019)\citenamefont
  {Mejuto-Zaera}, \citenamefont {Tubman},\ and\ \citenamefont
  {Whaley}}]{ASCIdmft}%
  \BibitemOpen
  \bibfield  {author} {\bibinfo {author} {\bibfnamefont {C.}~\bibnamefont
  {Mejuto-Zaera}}, \bibinfo {author} {\bibfnamefont {N.~M.}\ \bibnamefont
  {Tubman}},\ and\ \bibinfo {author} {\bibfnamefont {K.~B.}\ \bibnamefont
  {Whaley}},\ }\bibfield  {title} {\bibinfo {title} {Dynamical mean field
  theory simulations with the adaptive sampling configuration interaction
  method},\ }\href {https://doi.org/10.1103/PhysRevB.100.125165} {\bibfield
  {journal} {\bibinfo  {journal} {Phys. Rev. B}\ }\textbf {\bibinfo {volume}
  {100}},\ \bibinfo {pages} {125165} (\bibinfo {year} {2019})}\BibitemShut
  {NoStop}%
\bibitem [{\citenamefont {Wolf}\ \emph {et~al.}(2008)\citenamefont {Wolf},
  \citenamefont {Verstraete}, \citenamefont {Hastings},\ and\ \citenamefont
  {Cirac}}]{Wolf}%
  \BibitemOpen
  \bibfield  {author} {\bibinfo {author} {\bibfnamefont {M.~M.}\ \bibnamefont
  {Wolf}}, \bibinfo {author} {\bibfnamefont {F.}~\bibnamefont {Verstraete}},
  \bibinfo {author} {\bibfnamefont {M.~B.}\ \bibnamefont {Hastings}},\ and\
  \bibinfo {author} {\bibfnamefont {J.~I.}\ \bibnamefont {Cirac}},\ }\bibfield
  {title} {\bibinfo {title} {Area laws in quantum systems: Mutual information
  and correlations},\ }\href {https://doi.org/10.1103/PhysRevLett.100.070502}
  {\bibfield  {journal} {\bibinfo  {journal} {Phys. Rev. Lett.}\ }\textbf
  {\bibinfo {volume} {100}},\ \bibinfo {pages} {070502} (\bibinfo {year}
  {2008})}\BibitemShut {NoStop}%
\bibitem [{\citenamefont {Ding}\ \emph
  {et~al.}(2022{\natexlab{b}})\citenamefont {Ding}, \citenamefont {Zimboras},\
  and\ \citenamefont {Schilling}}]{Schilling_Entanglement_and_Locality_2023}%
  \BibitemOpen
  \bibfield  {author} {\bibinfo {author} {\bibfnamefont {L.}~\bibnamefont
  {Ding}}, \bibinfo {author} {\bibfnamefont {Z.}~\bibnamefont {Zimboras}},\
  and\ \bibinfo {author} {\bibfnamefont {C.}~\bibnamefont {Schilling}},\ }\href
  {https://doi.org/10.48550/ARXIV.2207.03377} {\bibinfo {title} {Quantifying
  electron entanglement faithfully}} (\bibinfo {year} {2022}{\natexlab{b}}),\
  \Eprint {https://arxiv.org/abs/2207.03377} {arXiv:2207.03377 [quant-ph]}
  \BibitemShut {NoStop}%
\bibitem [{\citenamefont {Ding}\ \emph {et~al.}(2023)\citenamefont {Ding},
  \citenamefont {Dünnweber},\ and\ \citenamefont
  {Schilling}}]{ding2023physical_entanglement}%
  \BibitemOpen
  \bibfield  {author} {\bibinfo {author} {\bibfnamefont {L.}~\bibnamefont
  {Ding}}, \bibinfo {author} {\bibfnamefont {G.}~\bibnamefont {Dünnweber}},\
  and\ \bibinfo {author} {\bibfnamefont {C.}~\bibnamefont {Schilling}},\
  }\bibfield  {title} {\bibinfo {title} {Physical entanglement between
  localized orbitals},\ }\href {https://doi.org/10.1088/2058-9565/ad00d9}
  {\bibfield  {journal} {\bibinfo  {journal} {Quantum Science and Technology}\
  }\textbf {\bibinfo {volume} {9}},\ \bibinfo {pages} {015005} (\bibinfo {year}
  {2023})}\BibitemShut {NoStop}%
\bibitem [{\citenamefont {Vedral}(2002)}]{Vedral_RMP}%
  \BibitemOpen
  \bibfield  {author} {\bibinfo {author} {\bibfnamefont {V.}~\bibnamefont
  {Vedral}},\ }\bibfield  {title} {\bibinfo {title} {The role of relative
  entropy in quantum information theory},\ }\href
  {https://doi.org/10.1103/RevModPhys.74.197} {\bibfield  {journal} {\bibinfo
  {journal} {Rev. Mod. Phys.}\ }\textbf {\bibinfo {volume} {74}},\ \bibinfo
  {pages} {197} (\bibinfo {year} {2002})}\BibitemShut {NoStop}%
\bibitem [{\citenamefont {Krueger}\ and\ \citenamefont
  {Werner}(2005)}]{krueger2005open}%
  \BibitemOpen
  \bibfield  {author} {\bibinfo {author} {\bibfnamefont {O.}~\bibnamefont
  {Krueger}}\ and\ \bibinfo {author} {\bibfnamefont {R.~F.}\ \bibnamefont
  {Werner}},\ }\href@noop {} {\bibinfo {title} {Some open problems in quantum
  information theory}} (\bibinfo {year} {2005}),\ \Eprint
  {https://arxiv.org/abs/quant-ph/0504166} {arXiv:quant-ph/0504166 [quant-ph]}
  \BibitemShut {NoStop}%
\bibitem [{\citenamefont {Wick}\ \emph {et~al.}(1952)\citenamefont {Wick},
  \citenamefont {Wightman},\ and\ \citenamefont {Wigner}}]{Wick1952}%
  \BibitemOpen
  \bibfield  {author} {\bibinfo {author} {\bibfnamefont {G.~C.}\ \bibnamefont
  {Wick}}, \bibinfo {author} {\bibfnamefont {A.~S.}\ \bibnamefont {Wightman}},\
  and\ \bibinfo {author} {\bibfnamefont {E.~P.}\ \bibnamefont {Wigner}},\
  }\bibfield  {title} {\bibinfo {title} {The intrinsic parity of elementary
  particles},\ }\href {https://doi.org/10.1103/PhysRev.88.101} {\bibfield
  {journal} {\bibinfo  {journal} {Phys. Rev.}\ }\textbf {\bibinfo {volume}
  {88}},\ \bibinfo {pages} {101} (\bibinfo {year} {1952})}\BibitemShut
  {NoStop}%
\bibitem [{\citenamefont {Wick}\ \emph {et~al.}(1970)\citenamefont {Wick},
  \citenamefont {Wightman},\ and\ \citenamefont {Wigner}}]{Wick1970}%
  \BibitemOpen
  \bibfield  {author} {\bibinfo {author} {\bibfnamefont {G.~C.}\ \bibnamefont
  {Wick}}, \bibinfo {author} {\bibfnamefont {A.~S.}\ \bibnamefont {Wightman}},\
  and\ \bibinfo {author} {\bibfnamefont {E.~P.}\ \bibnamefont {Wigner}},\
  }\bibfield  {title} {\bibinfo {title} {Superselection rule for charge},\
  }\href {https://doi.org/10.1103/PhysRevD.1.3267} {\bibfield  {journal}
  {\bibinfo  {journal} {Phys. Rev. D}\ }\textbf {\bibinfo {volume} {1}},\
  \bibinfo {pages} {3267} (\bibinfo {year} {1970})}\BibitemShut {NoStop}%
\bibitem [{\citenamefont {Bartlett}\ and\ \citenamefont
  {Wiseman}(2003)}]{Bartlett2003}%
  \BibitemOpen
  \bibfield  {author} {\bibinfo {author} {\bibfnamefont {S.~D.}\ \bibnamefont
  {Bartlett}}\ and\ \bibinfo {author} {\bibfnamefont {H.~M.}\ \bibnamefont
  {Wiseman}},\ }\bibfield  {title} {\bibinfo {title} {Entanglement constrained
  by superselection rules},\ }\href
  {https://doi.org/10.1103/PhysRevLett.91.097903} {\bibfield  {journal}
  {\bibinfo  {journal} {Phys. Rev. Lett.}\ }\textbf {\bibinfo {volume} {91}},\
  \bibinfo {pages} {097903} (\bibinfo {year} {2003})}\BibitemShut {NoStop}%
\bibitem [{\citenamefont {Ba\~nuls}\ \emph {et~al.}(2007)\citenamefont
  {Ba\~nuls}, \citenamefont {Cirac},\ and\ \citenamefont {Wolf}}]{Banhuls2007}%
  \BibitemOpen
  \bibfield  {author} {\bibinfo {author} {\bibfnamefont {M.-C.}\ \bibnamefont
  {Ba\~nuls}}, \bibinfo {author} {\bibfnamefont {J.~I.}\ \bibnamefont
  {Cirac}},\ and\ \bibinfo {author} {\bibfnamefont {M.~M.}\ \bibnamefont
  {Wolf}},\ }\bibfield  {title} {\bibinfo {title} {Entanglement in fermionic
  systems},\ }\href {https://doi.org/10.1103/PhysRevA.76.022311} {\bibfield
  {journal} {\bibinfo  {journal} {Phys. Rev. A}\ }\textbf {\bibinfo {volume}
  {76}},\ \bibinfo {pages} {022311} (\bibinfo {year} {2007})}\BibitemShut
  {NoStop}%
\bibitem [{\citenamefont {Friis}\ \emph {et~al.}(2013)\citenamefont {Friis},
  \citenamefont {Lee},\ and\ \citenamefont {Bruschi}}]{Friis2013}%
  \BibitemOpen
  \bibfield  {author} {\bibinfo {author} {\bibfnamefont {N.}~\bibnamefont
  {Friis}}, \bibinfo {author} {\bibfnamefont {A.~R.}\ \bibnamefont {Lee}},\
  and\ \bibinfo {author} {\bibfnamefont {D.~E.}\ \bibnamefont {Bruschi}},\
  }\bibfield  {title} {\bibinfo {title} {Fermionic-mode entanglement in quantum
  information},\ }\href {https://doi.org/10.1103/PhysRevA.87.022338} {\bibfield
   {journal} {\bibinfo  {journal} {Phys. Rev. A}\ }\textbf {\bibinfo {volume}
  {87}},\ \bibinfo {pages} {022338} (\bibinfo {year} {2013})}\BibitemShut
  {NoStop}%
\bibitem [{\citenamefont {Friis}(2016)}]{Friis2016}%
  \BibitemOpen
  \bibfield  {author} {\bibinfo {author} {\bibfnamefont {N.}~\bibnamefont
  {Friis}},\ }\bibfield  {title} {\bibinfo {title} {Reasonable fermionic
  quantum information theories require relativity},\ }\href
  {https://doi.org/10.1088/1367-2630/18/3/033014} {\bibfield  {journal}
  {\bibinfo  {journal} {New Journal of Physics}\ }\textbf {\bibinfo {volume}
  {18}},\ \bibinfo {pages} {033014} (\bibinfo {year} {2016})}\BibitemShut
  {NoStop}%
\bibitem [{\citenamefont {Benatti}\ \emph {et~al.}(2020)\citenamefont
  {Benatti}, \citenamefont {Floreanini}, \citenamefont {Franchini},\ and\
  \citenamefont {Marzolino}}]{Benatti2020}%
  \BibitemOpen
  \bibfield  {author} {\bibinfo {author} {\bibfnamefont {F.}~\bibnamefont
  {Benatti}}, \bibinfo {author} {\bibfnamefont {R.}~\bibnamefont {Floreanini}},
  \bibinfo {author} {\bibfnamefont {F.}~\bibnamefont {Franchini}},\ and\
  \bibinfo {author} {\bibfnamefont {U.}~\bibnamefont {Marzolino}},\ }\bibfield
  {title} {\bibinfo {title} {Entanglement in indistinguishable particle
  systems},\ }\href
  {https://doi.org/https://doi.org/10.1016/j.physrep.2020.07.003} {\bibfield
  {journal} {\bibinfo  {journal} {Physics Reports}\ }\textbf {\bibinfo {volume}
  {878}},\ \bibinfo {pages} {1} (\bibinfo {year} {2020})}\BibitemShut {NoStop}%
\bibitem [{\citenamefont {Vidal}\ \emph {et~al.}(2021)\citenamefont {Vidal},
  \citenamefont {Bera}, \citenamefont {Riera}, \citenamefont {Lewenstein},\
  and\ \citenamefont {Bera}}]{Vidal2021}%
  \BibitemOpen
  \bibfield  {author} {\bibinfo {author} {\bibfnamefont {N.~T.}\ \bibnamefont
  {Vidal}}, \bibinfo {author} {\bibfnamefont {M.~L.}\ \bibnamefont {Bera}},
  \bibinfo {author} {\bibfnamefont {A.}~\bibnamefont {Riera}}, \bibinfo
  {author} {\bibfnamefont {M.}~\bibnamefont {Lewenstein}},\ and\ \bibinfo
  {author} {\bibfnamefont {M.~N.}\ \bibnamefont {Bera}},\ }\bibfield  {title}
  {\bibinfo {title} {Quantum operations in an information theory for
  fermions},\ }\href {https://doi.org/10.1103/PhysRevA.104.032411} {\bibfield
  {journal} {\bibinfo  {journal} {Phys. Rev. A}\ }\textbf {\bibinfo {volume}
  {104}},\ \bibinfo {pages} {032411} (\bibinfo {year} {2021})}\BibitemShut
  {NoStop}%
\bibitem [{\citenamefont {Costa~de Almeida}\ and\ \citenamefont
  {Hauke}(2021)}]{RicardoCosta2021}%
  \BibitemOpen
  \bibfield  {author} {\bibinfo {author} {\bibfnamefont {R.}~\bibnamefont
  {Costa~de Almeida}}\ and\ \bibinfo {author} {\bibfnamefont {P.}~\bibnamefont
  {Hauke}},\ }\bibfield  {title} {\bibinfo {title} {From entanglement
  certification with quench dynamics to multipartite entanglement of
  interacting fermions},\ }\href
  {https://doi.org/10.1103/PhysRevResearch.3.L032051} {\bibfield  {journal}
  {\bibinfo  {journal} {Phys. Rev. Res.}\ }\textbf {\bibinfo {volume} {3}},\
  \bibinfo {pages} {L032051} (\bibinfo {year} {2021})}\BibitemShut {NoStop}%
\bibitem [{\citenamefont {Shen}\ \emph {et~al.}(2023)\citenamefont {Shen},
  \citenamefont {Barghathi}, \citenamefont {Maestro},\ and\ \citenamefont
  {Rubenstein}}]{DelMaestro_SSR}%
  \BibitemOpen
  \bibfield  {author} {\bibinfo {author} {\bibfnamefont {T.}~\bibnamefont
  {Shen}}, \bibinfo {author} {\bibfnamefont {H.}~\bibnamefont {Barghathi}},
  \bibinfo {author} {\bibfnamefont {A.~D.}\ \bibnamefont {Maestro}},\ and\
  \bibinfo {author} {\bibfnamefont {B.}~\bibnamefont {Rubenstein}},\
  }\href@noop {} {\bibinfo {title} {Disentangling the physics of the attractive
  {Hubbard} model via the accessible and symmetry-resolved entanglement
  entropies}} (\bibinfo {year} {2023}),\ \Eprint
  {https://arxiv.org/abs/2312.11746} {arXiv:2312.11746 [cond-mat.str-el]}
  \BibitemShut {NoStop}%
\bibitem [{\citenamefont {Ptaszy\ifmmode~\acute{n}\else \'{n}\fi{}ski}\ and\
  \citenamefont {Esposito}(2023)}]{Esposito2023}%
  \BibitemOpen
  \bibfield  {author} {\bibinfo {author} {\bibfnamefont {K.}~\bibnamefont
  {Ptaszy\ifmmode~\acute{n}\else \'{n}\fi{}ski}}\ and\ \bibinfo {author}
  {\bibfnamefont {M.}~\bibnamefont {Esposito}},\ }\bibfield  {title} {\bibinfo
  {title} {Fermionic one-body entanglement as a thermodynamic resource},\
  }\href {https://doi.org/10.1103/PhysRevLett.130.150201} {\bibfield  {journal}
  {\bibinfo  {journal} {Phys. Rev. Lett.}\ }\textbf {\bibinfo {volume} {130}},\
  \bibinfo {pages} {150201} (\bibinfo {year} {2023})}\BibitemShut {NoStop}%
\bibitem [{\citenamefont {Ding}\ \emph {et~al.}(2021)\citenamefont {Ding},
  \citenamefont {Mardazad}, \citenamefont {Das}, \citenamefont {Szalay},
  \citenamefont {Schollwöck}, \citenamefont {Zimborás},\ and\ \citenamefont
  {Schilling}}]{EntanglementVsCorrelation2021}%
  \BibitemOpen
  \bibfield  {author} {\bibinfo {author} {\bibfnamefont {L.}~\bibnamefont
  {Ding}}, \bibinfo {author} {\bibfnamefont {S.}~\bibnamefont {Mardazad}},
  \bibinfo {author} {\bibfnamefont {S.}~\bibnamefont {Das}}, \bibinfo {author}
  {\bibfnamefont {S.}~\bibnamefont {Szalay}}, \bibinfo {author} {\bibfnamefont
  {U.}~\bibnamefont {Schollwöck}}, \bibinfo {author} {\bibfnamefont
  {Z.}~\bibnamefont {Zimborás}},\ and\ \bibinfo {author} {\bibfnamefont
  {C.}~\bibnamefont {Schilling}},\ }\bibfield  {title} {\bibinfo {title}
  {Concept of orbital entanglement and correlation in quantum chemistry},\
  }\href {https://doi.org/10.1021/acs.jctc.0c00559} {\bibfield  {journal}
  {\bibinfo  {journal} {Journal of Chemical Theory and Computation}\ }\textbf
  {\bibinfo {volume} {17}},\ \bibinfo {pages} {79} (\bibinfo {year} {2021})},\
  \bibinfo {note} {pMID: 33430597}\BibitemShut {NoStop}%
\bibitem [{Note1()}]{Note1}%
  \BibitemOpen
  \bibinfo {note} {An unpublished version of the implemented cluster extension
  can be found at \protect \url
  {https://doi.org/10.5281/zenodo.10628157}}\BibitemShut {NoStop}%
\bibitem [{\citenamefont {Capone}\ \emph
  {et~al.}(2004{\natexlab{a}})\citenamefont {Capone}, \citenamefont {Civelli},
  \citenamefont {Kancharla}, \citenamefont {Castellani},\ and\ \citenamefont
  {Kotliar}}]{Capone2004PRB}%
  \BibitemOpen
  \bibfield  {author} {\bibinfo {author} {\bibfnamefont {M.}~\bibnamefont
  {Capone}}, \bibinfo {author} {\bibfnamefont {M.}~\bibnamefont {Civelli}},
  \bibinfo {author} {\bibfnamefont {S.~S.}\ \bibnamefont {Kancharla}}, \bibinfo
  {author} {\bibfnamefont {C.}~\bibnamefont {Castellani}},\ and\ \bibinfo
  {author} {\bibfnamefont {G.}~\bibnamefont {Kotliar}},\ }\bibfield  {title}
  {\bibinfo {title} {Cluster-dynamical mean-field theory of the density-driven
  {Mott} transition in the one-dimensional {Hubbard} model},\ }\href
  {https://doi.org/10.1103/PhysRevB.69.195105} {\bibfield  {journal} {\bibinfo
  {journal} {Phys. Rev. B}\ }\textbf {\bibinfo {volume} {69}},\ \bibinfo
  {pages} {195105} (\bibinfo {year} {2004}{\natexlab{a}})}\BibitemShut
  {NoStop}%
\bibitem [{\citenamefont {Koch}\ \emph {et~al.}(2008)\citenamefont {Koch},
  \citenamefont {Sangiovanni},\ and\ \citenamefont {Gunnarsson}}]{Koch2008PRB}%
  \BibitemOpen
  \bibfield  {author} {\bibinfo {author} {\bibfnamefont {E.}~\bibnamefont
  {Koch}}, \bibinfo {author} {\bibfnamefont {G.}~\bibnamefont {Sangiovanni}},\
  and\ \bibinfo {author} {\bibfnamefont {O.}~\bibnamefont {Gunnarsson}},\
  }\bibfield  {title} {\bibinfo {title} {Sum rules and bath parametrization for
  quantum cluster theories},\ }\href
  {https://doi.org/10.1103/PhysRevB.78.115102} {\bibfield  {journal} {\bibinfo
  {journal} {Phys. Rev. B}\ }\textbf {\bibinfo {volume} {78}},\ \bibinfo
  {pages} {115102} (\bibinfo {year} {2008})}\BibitemShut {NoStop}%
\bibitem [{\citenamefont {Mejuto-Zaera}\ \emph {et~al.}(2021)\citenamefont
  {Mejuto-Zaera}, \citenamefont {Weng}, \citenamefont {Romanova}, \citenamefont
  {Cotton}, \citenamefont {Whaley}, \citenamefont {Tubman},\ and\ \citenamefont
  {Vl{\v{c}}ek}}]{MejutoZaera2021}%
  \BibitemOpen
  \bibfield  {author} {\bibinfo {author} {\bibfnamefont {C.}~\bibnamefont
  {Mejuto-Zaera}}, \bibinfo {author} {\bibfnamefont {G.}~\bibnamefont {Weng}},
  \bibinfo {author} {\bibfnamefont {M.}~\bibnamefont {Romanova}}, \bibinfo
  {author} {\bibfnamefont {S.~J.}\ \bibnamefont {Cotton}}, \bibinfo {author}
  {\bibfnamefont {K.~B.}\ \bibnamefont {Whaley}}, \bibinfo {author}
  {\bibfnamefont {N.~M.}\ \bibnamefont {Tubman}},\ and\ \bibinfo {author}
  {\bibfnamefont {V.}~\bibnamefont {Vl{\v{c}}ek}},\ }\bibfield  {title}
  {\bibinfo {title} {Are multi-quasiparticle interactions important in
  molecular ionization?},\ }\href {https://doi.org/10.1063/5.0044060}
  {\bibfield  {journal} {\bibinfo  {journal} {The Journal of Chemical Physics}\
  }\textbf {\bibinfo {volume} {154}},\ \bibinfo {pages} {121101} (\bibinfo
  {year} {2021})}\BibitemShut {NoStop}%
\bibitem [{\citenamefont {Levine}\ \emph {et~al.}(2020)\citenamefont {Levine},
  \citenamefont {Hait}, \citenamefont {Tubman}, \citenamefont {Lehtola},
  \citenamefont {Whaley},\ and\ \citenamefont {Head-Gordon}}]{Levine2020}%
  \BibitemOpen
  \bibfield  {author} {\bibinfo {author} {\bibfnamefont {D.~S.}\ \bibnamefont
  {Levine}}, \bibinfo {author} {\bibfnamefont {D.}~\bibnamefont {Hait}},
  \bibinfo {author} {\bibfnamefont {N.~M.}\ \bibnamefont {Tubman}}, \bibinfo
  {author} {\bibfnamefont {S.}~\bibnamefont {Lehtola}}, \bibinfo {author}
  {\bibfnamefont {K.~B.}\ \bibnamefont {Whaley}},\ and\ \bibinfo {author}
  {\bibfnamefont {M.}~\bibnamefont {Head-Gordon}},\ }\bibfield  {title}
  {\bibinfo {title} {{CASSCF} with extremely large active spaces using the
  adaptive sampling configuration interaction method},\ }\href
  {https://doi.org/10.1021/acs.jctc.9b01255} {\bibfield  {journal} {\bibinfo
  {journal} {Journal of Chemical Theory and Computation}\ }\textbf {\bibinfo
  {volume} {16}},\ \bibinfo {pages} {2340} (\bibinfo {year}
  {2020})}\BibitemShut {NoStop}%
\bibitem [{\citenamefont {Mejuto-Zaera}\ \emph {et~al.}(2022)\citenamefont
  {Mejuto-Zaera}, \citenamefont {Tzeli}, \citenamefont {Williams-Young},
  \citenamefont {Tubman}, \citenamefont {Matoušek}, \citenamefont {Brabec},
  \citenamefont {Veis}, \citenamefont {Xantheas},\ and\ \citenamefont
  {de~Jong}}]{Mejuto2022}%
  \BibitemOpen
  \bibfield  {author} {\bibinfo {author} {\bibfnamefont {C.}~\bibnamefont
  {Mejuto-Zaera}}, \bibinfo {author} {\bibfnamefont {D.}~\bibnamefont {Tzeli}},
  \bibinfo {author} {\bibfnamefont {D.}~\bibnamefont {Williams-Young}},
  \bibinfo {author} {\bibfnamefont {N.~M.}\ \bibnamefont {Tubman}}, \bibinfo
  {author} {\bibfnamefont {M.}~\bibnamefont {Matoušek}}, \bibinfo {author}
  {\bibfnamefont {J.}~\bibnamefont {Brabec}}, \bibinfo {author} {\bibfnamefont
  {L.}~\bibnamefont {Veis}}, \bibinfo {author} {\bibfnamefont {S.~S.}\
  \bibnamefont {Xantheas}},\ and\ \bibinfo {author} {\bibfnamefont {W.~A.}\
  \bibnamefont {de~Jong}},\ }\bibfield  {title} {\bibinfo {title} {The effect
  of geometry, spin, and orbital optimization in achieving accurate, correlated
  results for iron–sulfur cubanes},\ }\href
  {https://doi.org/10.1021/acs.jctc.1c00830} {\bibfield  {journal} {\bibinfo
  {journal} {Journal of Chemical Theory and Computation}\ }\textbf {\bibinfo
  {volume} {18}},\ \bibinfo {pages} {687} (\bibinfo {year} {2022})}\BibitemShut
  {NoStop}%
\bibitem [{\citenamefont {Tubman}\ \emph {et~al.}(2016)\citenamefont {Tubman},
  \citenamefont {Lee}, \citenamefont {Takeshita}, \citenamefont {Head-Gordon},\
  and\ \citenamefont {Whaley}}]{Tubman2016}%
  \BibitemOpen
  \bibfield  {author} {\bibinfo {author} {\bibfnamefont {N.~M.}\ \bibnamefont
  {Tubman}}, \bibinfo {author} {\bibfnamefont {J.}~\bibnamefont {Lee}},
  \bibinfo {author} {\bibfnamefont {T.~Y.}\ \bibnamefont {Takeshita}}, \bibinfo
  {author} {\bibfnamefont {M.}~\bibnamefont {Head-Gordon}},\ and\ \bibinfo
  {author} {\bibfnamefont {K.~B.}\ \bibnamefont {Whaley}},\ }\bibfield  {title}
  {\bibinfo {title} {A deterministic alternative to the full configuration
  interaction quantum monte carlo method},\ }\href
  {https://doi.org/10.1063/1.4955109} {\bibfield  {journal} {\bibinfo
  {journal} {The Journal of Chemical Physics}\ }\textbf {\bibinfo {volume}
  {145}},\ \bibinfo {pages} {044112} (\bibinfo {year} {2016})}\BibitemShut
  {NoStop}%
\bibitem [{\citenamefont {Williams-Young}\ \emph {et~al.}(2023)\citenamefont
  {Williams-Young}, \citenamefont {Tubman}, \citenamefont {Mejuto-Zaera},\ and\
  \citenamefont {de~Jong}}]{WilliamsYoung2023}%
  \BibitemOpen
  \bibfield  {author} {\bibinfo {author} {\bibfnamefont {D.~B.}\ \bibnamefont
  {Williams-Young}}, \bibinfo {author} {\bibfnamefont {N.~M.}\ \bibnamefont
  {Tubman}}, \bibinfo {author} {\bibfnamefont {C.}~\bibnamefont
  {Mejuto-Zaera}},\ and\ \bibinfo {author} {\bibfnamefont {W.~A.}\ \bibnamefont
  {de~Jong}},\ }\bibfield  {title} {\bibinfo {title} {{A parallel, distributed
  memory implementation of the adaptive sampling configuration interaction
  method}},\ }\href {https://doi.org/10.1063/5.0148650} {\bibfield  {journal}
  {\bibinfo  {journal} {The Journal of Chemical Physics}\ }\textbf {\bibinfo
  {volume} {158}},\ \bibinfo {pages} {214109} (\bibinfo {year}
  {2023})}\BibitemShut {NoStop}%
\bibitem [{\citenamefont {Bravyi}\ and\ \citenamefont
  {Gosset}(2017)}]{Bravyi2016}%
  \BibitemOpen
  \bibfield  {author} {\bibinfo {author} {\bibfnamefont {S.}~\bibnamefont
  {Bravyi}}\ and\ \bibinfo {author} {\bibfnamefont {D.}~\bibnamefont
  {Gosset}},\ }\bibfield  {title} {\bibinfo {title} {Complexity of quantum
  impurity problems},\ }\href {https://doi.org/10.1007/s00220-017-2976-9}
  {\bibfield  {journal} {\bibinfo  {journal} {Communications in Mathematical
  Physics}\ }\textbf {\bibinfo {volume} {356}},\ \bibinfo {pages} {451}
  (\bibinfo {year} {2017})}\BibitemShut {NoStop}%
\bibitem [{\citenamefont {Mejuto-Zaera}\ \emph {et~al.}(2020)\citenamefont
  {Mejuto-Zaera}, \citenamefont {Zepeda-N\'u\~nez}, \citenamefont {Lindsey},
  \citenamefont {Tubman}, \citenamefont {Whaley},\ and\ \citenamefont
  {Lin}}]{ASCIfit}%
  \BibitemOpen
  \bibfield  {author} {\bibinfo {author} {\bibfnamefont {C.}~\bibnamefont
  {Mejuto-Zaera}}, \bibinfo {author} {\bibfnamefont {L.}~\bibnamefont
  {Zepeda-N\'u\~nez}}, \bibinfo {author} {\bibfnamefont {M.}~\bibnamefont
  {Lindsey}}, \bibinfo {author} {\bibfnamefont {N.}~\bibnamefont {Tubman}},
  \bibinfo {author} {\bibfnamefont {B.}~\bibnamefont {Whaley}},\ and\ \bibinfo
  {author} {\bibfnamefont {L.}~\bibnamefont {Lin}},\ }\bibfield  {title}
  {\bibinfo {title} {Efficient hybridization fitting for dynamical mean-field
  theory via semi-definite relaxation},\ }\href
  {https://doi.org/10.1103/PhysRevB.101.035143} {\bibfield  {journal} {\bibinfo
   {journal} {Phys. Rev. B}\ }\textbf {\bibinfo {volume} {101}},\ \bibinfo
  {pages} {035143} (\bibinfo {year} {2020})}\BibitemShut {NoStop}%
\bibitem [{\citenamefont {Roósz}\ \emph {et~al.}(2023)\citenamefont {Roósz},
  \citenamefont {Kauch}, \citenamefont {Bippus}, \citenamefont {Wieser},\ and\
  \citenamefont {Held}}]{Held_2RDMfrom2GF}%
  \BibitemOpen
  \bibfield  {author} {\bibinfo {author} {\bibfnamefont {G.}~\bibnamefont
  {Roósz}}, \bibinfo {author} {\bibfnamefont {A.}~\bibnamefont {Kauch}},
  \bibinfo {author} {\bibfnamefont {F.}~\bibnamefont {Bippus}}, \bibinfo
  {author} {\bibfnamefont {D.}~\bibnamefont {Wieser}},\ and\ \bibinfo {author}
  {\bibfnamefont {K.}~\bibnamefont {Held}},\ }\href@noop {} {\bibinfo {title}
  {Two-site reduced density matrix from one- and two-particle {Green}'s
  functions}} (\bibinfo {year} {2023}),\ \Eprint
  {https://arxiv.org/abs/2312.14275} {arXiv:2312.14275 [cond-mat.str-el]}
  \BibitemShut {NoStop}%
\bibitem [{\citenamefont {Stocker}\ \emph {et~al.}(2022)\citenamefont
  {Stocker}, \citenamefont {Sack}, \citenamefont {Ferguson},\ and\
  \citenamefont {Zilberberg}}]{LidiaStocker2022}%
  \BibitemOpen
  \bibfield  {author} {\bibinfo {author} {\bibfnamefont {L.}~\bibnamefont
  {Stocker}}, \bibinfo {author} {\bibfnamefont {S.~H.}\ \bibnamefont {Sack}},
  \bibinfo {author} {\bibfnamefont {M.~S.}\ \bibnamefont {Ferguson}},\ and\
  \bibinfo {author} {\bibfnamefont {O.}~\bibnamefont {Zilberberg}},\ }\bibfield
   {title} {\bibinfo {title} {Entanglement-based observables for quantum
  impurities},\ }\href {https://doi.org/10.1103/PhysRevResearch.4.043177}
  {\bibfield  {journal} {\bibinfo  {journal} {Phys. Rev. Res.}\ }\textbf
  {\bibinfo {volume} {4}},\ \bibinfo {pages} {043177} (\bibinfo {year}
  {2022})}\BibitemShut {NoStop}%
\bibitem [{\citenamefont {Carisch}\ and\ \citenamefont
  {Zilberberg}(2023)}]{Carisch2023}%
  \BibitemOpen
  \bibfield  {author} {\bibinfo {author} {\bibfnamefont {C.}~\bibnamefont
  {Carisch}}\ and\ \bibinfo {author} {\bibfnamefont {O.}~\bibnamefont
  {Zilberberg}},\ }\bibfield  {title} {\bibinfo {title} {Efficient separation
  of quantum from classical correlations for mixed states with a fixed
  charge},\ }\href {https://doi.org/10.22331/q-2023-03-20-954} {\bibfield
  {journal} {\bibinfo  {journal} {{Quantum}}\ }\textbf {\bibinfo {volume}
  {7}},\ \bibinfo {pages} {954} (\bibinfo {year} {2023})}\BibitemShut {NoStop}%
\bibitem [{\citenamefont {Lieb}\ and\ \citenamefont {Ruskai}(1973)}]{Lieb_SSA}%
  \BibitemOpen
  \bibfield  {author} {\bibinfo {author} {\bibfnamefont {E.~H.}\ \bibnamefont
  {Lieb}}\ and\ \bibinfo {author} {\bibfnamefont {M.~B.}\ \bibnamefont
  {Ruskai}},\ }\bibfield  {title} {\bibinfo {title} {A fundamental property of
  quantum-mechanical entropy},\ }\href
  {https://doi.org/10.1103/PhysRevLett.30.434} {\bibfield  {journal} {\bibinfo
  {journal} {Phys. Rev. Lett.}\ }\textbf {\bibinfo {volume} {30}},\ \bibinfo
  {pages} {434} (\bibinfo {year} {1973})}\BibitemShut {NoStop}%
\bibitem [{\citenamefont {Tajik}\ \emph {et~al.}(2023)\citenamefont {Tajik},
  \citenamefont {Kukuljan}, \citenamefont {Sotiriadis}, \citenamefont {Rauer},
  \citenamefont {Schweigler}, \citenamefont {Cataldini}, \citenamefont
  {Sabino}, \citenamefont {Moller}, \citenamefont {Sch\"{u}ttelkopf},
  \citenamefont {Ji}, \citenamefont {Sels}, \citenamefont {Demler},\ and\
  \citenamefont {Schmiedmayer}}]{Tajik2023}%
  \BibitemOpen
  \bibfield  {author} {\bibinfo {author} {\bibfnamefont {M.}~\bibnamefont
  {Tajik}}, \bibinfo {author} {\bibfnamefont {I.}~\bibnamefont {Kukuljan}},
  \bibinfo {author} {\bibfnamefont {S.}~\bibnamefont {Sotiriadis}}, \bibinfo
  {author} {\bibfnamefont {B.}~\bibnamefont {Rauer}}, \bibinfo {author}
  {\bibfnamefont {T.}~\bibnamefont {Schweigler}}, \bibinfo {author}
  {\bibfnamefont {F.}~\bibnamefont {Cataldini}}, \bibinfo {author}
  {\bibfnamefont {J.}~\bibnamefont {Sabino}}, \bibinfo {author} {\bibfnamefont
  {F.}~\bibnamefont {Moller}}, \bibinfo {author} {\bibfnamefont
  {P.}~\bibnamefont {Sch\"{u}ttelkopf}}, \bibinfo {author} {\bibfnamefont
  {S.-C.}\ \bibnamefont {Ji}}, \bibinfo {author} {\bibfnamefont
  {D.}~\bibnamefont {Sels}}, \bibinfo {author} {\bibfnamefont {E.}~\bibnamefont
  {Demler}},\ and\ \bibinfo {author} {\bibfnamefont {J.}~\bibnamefont
  {Schmiedmayer}},\ }\bibfield  {title} {\bibinfo {title} {Verification of the
  area law of mutual information in a quantum field simulator},\ }\href
  {https://doi.org/10.1038/s41567-023-02027-1} {\bibfield  {journal} {\bibinfo
  {journal} {Nature Physics}\ }\textbf {\bibinfo {volume} {19}},\ \bibinfo
  {pages} {1022} (\bibinfo {year} {2023})}\BibitemShut {NoStop}%
\bibitem [{\citenamefont {Sch\"afer}\ \emph {et~al.}(2021)\citenamefont
  {Sch\"afer}, \citenamefont {Wentzell}, \citenamefont {\ifmmode~\check{S}\else
  \v{S}\fi{}imkovic}, \citenamefont {He}, \citenamefont {Hille}, \citenamefont
  {Klett}, \citenamefont {Eckhardt}, \citenamefont {Arzhang}, \citenamefont
  {Harkov}, \citenamefont {Le~R\'egent}, \citenamefont {Kirsch}, \citenamefont
  {Wang}, \citenamefont {Kim}, \citenamefont {Kozik}, \citenamefont {Stepanov},
  \citenamefont {Kauch}, \citenamefont {Andergassen}, \citenamefont {Hansmann},
  \citenamefont {Rohe}, \citenamefont {Vilk}, \citenamefont {LeBlanc},
  \citenamefont {Zhang}, \citenamefont {Tremblay}, \citenamefont {Ferrero},
  \citenamefont {Parcollet},\ and\ \citenamefont {Georges}}]{Schafer21Multi}%
  \BibitemOpen
  \bibfield  {author} {\bibinfo {author} {\bibfnamefont {T.}~\bibnamefont
  {Sch\"afer}}, \bibinfo {author} {\bibfnamefont {N.}~\bibnamefont {Wentzell}},
  \bibinfo {author} {\bibfnamefont {F.}~\bibnamefont {\ifmmode~\check{S}\else
  \v{S}\fi{}imkovic}}, \bibinfo {author} {\bibfnamefont {Y.-Y.}\ \bibnamefont
  {He}}, \bibinfo {author} {\bibfnamefont {C.}~\bibnamefont {Hille}}, \bibinfo
  {author} {\bibfnamefont {M.}~\bibnamefont {Klett}}, \bibinfo {author}
  {\bibfnamefont {C.~J.}\ \bibnamefont {Eckhardt}}, \bibinfo {author}
  {\bibfnamefont {B.}~\bibnamefont {Arzhang}}, \bibinfo {author} {\bibfnamefont
  {V.}~\bibnamefont {Harkov}}, \bibinfo {author} {\bibfnamefont {F.~m. c.-M.}\
  \bibnamefont {Le~R\'egent}}, \bibinfo {author} {\bibfnamefont
  {A.}~\bibnamefont {Kirsch}}, \bibinfo {author} {\bibfnamefont
  {Y.}~\bibnamefont {Wang}}, \bibinfo {author} {\bibfnamefont {A.~J.}\
  \bibnamefont {Kim}}, \bibinfo {author} {\bibfnamefont {E.}~\bibnamefont
  {Kozik}}, \bibinfo {author} {\bibfnamefont {E.~A.}\ \bibnamefont {Stepanov}},
  \bibinfo {author} {\bibfnamefont {A.}~\bibnamefont {Kauch}}, \bibinfo
  {author} {\bibfnamefont {S.}~\bibnamefont {Andergassen}}, \bibinfo {author}
  {\bibfnamefont {P.}~\bibnamefont {Hansmann}}, \bibinfo {author}
  {\bibfnamefont {D.}~\bibnamefont {Rohe}}, \bibinfo {author} {\bibfnamefont
  {Y.~M.}\ \bibnamefont {Vilk}}, \bibinfo {author} {\bibfnamefont {J.~P.~F.}\
  \bibnamefont {LeBlanc}}, \bibinfo {author} {\bibfnamefont {S.}~\bibnamefont
  {Zhang}}, \bibinfo {author} {\bibfnamefont {A.-M.~S.}\ \bibnamefont
  {Tremblay}}, \bibinfo {author} {\bibfnamefont {M.}~\bibnamefont {Ferrero}},
  \bibinfo {author} {\bibfnamefont {O.}~\bibnamefont {Parcollet}},\ and\
  \bibinfo {author} {\bibfnamefont {A.}~\bibnamefont {Georges}},\ }\bibfield
  {title} {\bibinfo {title} {Tracking the footprints of spin fluctuations: A
  multimethod, multimessenger study of the two-dimensional hubbard model},\
  }\href {https://doi.org/10.1103/PhysRevX.11.011058} {\bibfield  {journal}
  {\bibinfo  {journal} {Phys. Rev. X}\ }\textbf {\bibinfo {volume} {11}},\
  \bibinfo {pages} {011058} (\bibinfo {year} {2021})}\BibitemShut {NoStop}%
\bibitem [{\citenamefont {Meixner}\ \emph {et~al.}(2024)\citenamefont
  {Meixner}, \citenamefont {Menke}, \citenamefont {Klett}, \citenamefont
  {Heinzelmann}, \citenamefont {Andergassen}, \citenamefont {Hansmann},\ and\
  \citenamefont {Schäfer}}]{Schafer24Mott}%
  \BibitemOpen
  \bibfield  {author} {\bibinfo {author} {\bibfnamefont {M.}~\bibnamefont
  {Meixner}}, \bibinfo {author} {\bibfnamefont {H.}~\bibnamefont {Menke}},
  \bibinfo {author} {\bibfnamefont {M.}~\bibnamefont {Klett}}, \bibinfo
  {author} {\bibfnamefont {S.}~\bibnamefont {Heinzelmann}}, \bibinfo {author}
  {\bibfnamefont {S.}~\bibnamefont {Andergassen}}, \bibinfo {author}
  {\bibfnamefont {P.}~\bibnamefont {Hansmann}},\ and\ \bibinfo {author}
  {\bibfnamefont {T.}~\bibnamefont {Schäfer}},\ }\href@noop {} {\bibinfo
  {title} {Mott transition and pseudogap of the square-lattice hubbard model:
  results from center-focused cellular dynamical mean-field theory}} (\bibinfo
  {year} {2024}),\ \Eprint {https://arxiv.org/abs/2310.17302} {arXiv:2310.17302
  [cond-mat.str-el]} \BibitemShut {NoStop}%
\bibitem [{\citenamefont {Zheng}\ \emph {et~al.}(2017)\citenamefont {Zheng},
  \citenamefont {Chung}, \citenamefont {Corboz}, \citenamefont {Ehlers},
  \citenamefont {Qin}, \citenamefont {Noack}, \citenamefont {Shi},
  \citenamefont {White}, \citenamefont {Zhang},\ and\ \citenamefont
  {Chan}}]{Shiwei2017}%
  \BibitemOpen
  \bibfield  {author} {\bibinfo {author} {\bibfnamefont {B.-X.}\ \bibnamefont
  {Zheng}}, \bibinfo {author} {\bibfnamefont {C.-M.}\ \bibnamefont {Chung}},
  \bibinfo {author} {\bibfnamefont {P.}~\bibnamefont {Corboz}}, \bibinfo
  {author} {\bibfnamefont {G.}~\bibnamefont {Ehlers}}, \bibinfo {author}
  {\bibfnamefont {M.-P.}\ \bibnamefont {Qin}}, \bibinfo {author} {\bibfnamefont
  {R.~M.}\ \bibnamefont {Noack}}, \bibinfo {author} {\bibfnamefont
  {H.}~\bibnamefont {Shi}}, \bibinfo {author} {\bibfnamefont {S.~R.}\
  \bibnamefont {White}}, \bibinfo {author} {\bibfnamefont {S.}~\bibnamefont
  {Zhang}},\ and\ \bibinfo {author} {\bibfnamefont {G.~K.-L.}\ \bibnamefont
  {Chan}},\ }\bibfield  {title} {\bibinfo {title} {Stripe order in the
  underdoped region of the two-dimensional hubbard model},\ }\href
  {https://doi.org/10.1126/science.aam7127} {\bibfield  {journal} {\bibinfo
  {journal} {Science}\ }\textbf {\bibinfo {volume} {358}},\ \bibinfo {pages}
  {1155–1160} (\bibinfo {year} {2017})}\BibitemShut {NoStop}%
\bibitem [{\citenamefont {Wietek}\ \emph {et~al.}(2021)\citenamefont {Wietek},
  \citenamefont {Rossi}, \citenamefont {\ifmmode~\check{S}\else
  \v{S}\fi{}imkovic}, \citenamefont {Klett}, \citenamefont {Hansmann},
  \citenamefont {Ferrero}, \citenamefont {Stoudenmire}, \citenamefont
  {Sch\"afer},\ and\ \citenamefont {Georges}}]{MEETS2021}%
  \BibitemOpen
  \bibfield  {author} {\bibinfo {author} {\bibfnamefont {A.}~\bibnamefont
  {Wietek}}, \bibinfo {author} {\bibfnamefont {R.}~\bibnamefont {Rossi}},
  \bibinfo {author} {\bibfnamefont {F.}~\bibnamefont {\ifmmode~\check{S}\else
  \v{S}\fi{}imkovic}}, \bibinfo {author} {\bibfnamefont {M.}~\bibnamefont
  {Klett}}, \bibinfo {author} {\bibfnamefont {P.}~\bibnamefont {Hansmann}},
  \bibinfo {author} {\bibfnamefont {M.}~\bibnamefont {Ferrero}}, \bibinfo
  {author} {\bibfnamefont {E.~M.}\ \bibnamefont {Stoudenmire}}, \bibinfo
  {author} {\bibfnamefont {T.}~\bibnamefont {Sch\"afer}},\ and\ \bibinfo
  {author} {\bibfnamefont {A.}~\bibnamefont {Georges}},\ }\bibfield  {title}
  {\bibinfo {title} {Mott insulating states with competing orders in the
  triangular lattice hubbard model},\ }\href
  {https://doi.org/10.1103/PhysRevX.11.041013} {\bibfield  {journal} {\bibinfo
  {journal} {Phys. Rev. X}\ }\textbf {\bibinfo {volume} {11}},\ \bibinfo
  {pages} {041013} (\bibinfo {year} {2021})}\BibitemShut {NoStop}%
\bibitem [{\citenamefont {Xu}\ \emph {et~al.}(2023)\citenamefont {Xu},
  \citenamefont {Chung}, \citenamefont {Qin}, \citenamefont {Schollwöck},
  \citenamefont {White},\ and\ \citenamefont {Zhang}}]{Shiwei2023}%
  \BibitemOpen
  \bibfield  {author} {\bibinfo {author} {\bibfnamefont {H.}~\bibnamefont
  {Xu}}, \bibinfo {author} {\bibfnamefont {C.-M.}\ \bibnamefont {Chung}},
  \bibinfo {author} {\bibfnamefont {M.}~\bibnamefont {Qin}}, \bibinfo {author}
  {\bibfnamefont {U.}~\bibnamefont {Schollwöck}}, \bibinfo {author}
  {\bibfnamefont {S.~R.}\ \bibnamefont {White}},\ and\ \bibinfo {author}
  {\bibfnamefont {S.}~\bibnamefont {Zhang}},\ }\href@noop {} {\bibinfo {title}
  {Coexistence of superconductivity with partially filled stripes in the
  hubbard model}} (\bibinfo {year} {2023}),\ \Eprint
  {https://arxiv.org/abs/2303.08376} {arXiv:2303.08376 [cond-mat.supr-con]}
  \BibitemShut {NoStop}%
\bibitem [{\citenamefont {Capone}\ \emph
  {et~al.}(2004{\natexlab{b}})\citenamefont {Capone}, \citenamefont {Fabrizio},
  \citenamefont {Castellani},\ and\ \citenamefont {Tosatti}}]{Capone2004PRL}%
  \BibitemOpen
  \bibfield  {author} {\bibinfo {author} {\bibfnamefont {M.}~\bibnamefont
  {Capone}}, \bibinfo {author} {\bibfnamefont {M.}~\bibnamefont {Fabrizio}},
  \bibinfo {author} {\bibfnamefont {C.}~\bibnamefont {Castellani}},\ and\
  \bibinfo {author} {\bibfnamefont {E.}~\bibnamefont {Tosatti}},\ }\bibfield
  {title} {\bibinfo {title} {Strongly correlated superconductivity and
  pseudogap phase near a multiband {Mott} insulator},\ }\href
  {https://doi.org/10.1103/PhysRevLett.93.047001} {\bibfield  {journal}
  {\bibinfo  {journal} {Phys. Rev. Lett.}\ }\textbf {\bibinfo {volume} {93}},\
  \bibinfo {pages} {047001} (\bibinfo {year} {2004}{\natexlab{b}})}\BibitemShut
  {NoStop}%
\bibitem [{\citenamefont {de' Medici}\ \emph {et~al.}(2014)\citenamefont {de'
  Medici}, \citenamefont {Giovannetti},\ and\ \citenamefont
  {Capone}}]{Medici2014PRL}%
  \BibitemOpen
  \bibfield  {author} {\bibinfo {author} {\bibfnamefont {L.}~\bibnamefont {de'
  Medici}}, \bibinfo {author} {\bibfnamefont {G.}~\bibnamefont {Giovannetti}},\
  and\ \bibinfo {author} {\bibfnamefont {M.}~\bibnamefont {Capone}},\
  }\bibfield  {title} {\bibinfo {title} {Selective {Mott} physics as a key to
  iron superconductors},\ }\href
  {https://doi.org/10.1103/PhysRevLett.112.177001} {\bibfield  {journal}
  {\bibinfo  {journal} {Phys. Rev. Lett.}\ }\textbf {\bibinfo {volume} {112}},\
  \bibinfo {pages} {177001} (\bibinfo {year} {2014})}\BibitemShut {NoStop}%
\bibitem [{\citenamefont {Werner}\ \emph {et~al.}(2016)\citenamefont {Werner},
  \citenamefont {Hoshino},\ and\ \citenamefont {Shinaoka}}]{Werner2016PRB}%
  \BibitemOpen
  \bibfield  {author} {\bibinfo {author} {\bibfnamefont {P.}~\bibnamefont
  {Werner}}, \bibinfo {author} {\bibfnamefont {S.}~\bibnamefont {Hoshino}},\
  and\ \bibinfo {author} {\bibfnamefont {H.}~\bibnamefont {Shinaoka}},\
  }\bibfield  {title} {\bibinfo {title} {Spin-freezing perspective on
  cuprates},\ }\href {https://doi.org/10.1103/PhysRevB.94.245134} {\bibfield
  {journal} {\bibinfo  {journal} {Phys. Rev. B}\ }\textbf {\bibinfo {volume}
  {94}},\ \bibinfo {pages} {245134} (\bibinfo {year} {2016})}\BibitemShut
  {NoStop}%
\bibitem [{\citenamefont {Tomczak}\ \emph {et~al.}(2012)\citenamefont
  {Tomczak}, \citenamefont {{Van Schilfgaarde}},\ and\ \citenamefont
  {Kotliar}}]{Tomczak2012PRL}%
  \BibitemOpen
  \bibfield  {author} {\bibinfo {author} {\bibfnamefont {J.~M.}\ \bibnamefont
  {Tomczak}}, \bibinfo {author} {\bibfnamefont {M.}~\bibnamefont {{Van
  Schilfgaarde}}},\ and\ \bibinfo {author} {\bibfnamefont {G.}~\bibnamefont
  {Kotliar}},\ }\bibfield  {title} {\bibinfo {title} {{Many-body effects in
  iron pnictides and chalcogenides: Nonlocal versus dynamic origin of effective
  masses}},\ }\href {https://doi.org/10.1103/PhysRevLett.109.237010} {\bibfield
   {journal} {\bibinfo  {journal} {Phys. Rev. Lett.}\ }\textbf {\bibinfo
  {volume} {109}},\ \bibinfo {pages} {237010} (\bibinfo {year}
  {2012})}\BibitemShut {NoStop}%
\bibitem [{\citenamefont {Amaricci}\ \emph {et~al.}(2015)\citenamefont
  {Amaricci}, \citenamefont {Budich}, \citenamefont {Capone}, \citenamefont
  {Trauzettel},\ and\ \citenamefont {Sangiovanni}}]{AA_BHZ_PRL}%
  \BibitemOpen
  \bibfield  {author} {\bibinfo {author} {\bibfnamefont {A.}~\bibnamefont
  {Amaricci}}, \bibinfo {author} {\bibfnamefont {J.~C.}\ \bibnamefont
  {Budich}}, \bibinfo {author} {\bibfnamefont {M.}~\bibnamefont {Capone}},
  \bibinfo {author} {\bibfnamefont {B.}~\bibnamefont {Trauzettel}},\ and\
  \bibinfo {author} {\bibfnamefont {G.}~\bibnamefont {Sangiovanni}},\
  }\bibfield  {title} {\bibinfo {title} {First-order character and observable
  signatures of topological quantum phase transitions},\ }\href
  {https://doi.org/10.1103/PhysRevLett.114.185701} {\bibfield  {journal}
  {\bibinfo  {journal} {Phys. Rev. Lett.}\ }\textbf {\bibinfo {volume} {114}},\
  \bibinfo {pages} {185701} (\bibinfo {year} {2015})}\BibitemShut {NoStop}%
\bibitem [{\citenamefont {Crippa}\ \emph {et~al.}(2021)\citenamefont {Crippa},
  \citenamefont {Amaricci}, \citenamefont {Adler}, \citenamefont
  {Sangiovanni},\ and\ \citenamefont {Capone}}]{CrippaBHZ}%
  \BibitemOpen
  \bibfield  {author} {\bibinfo {author} {\bibfnamefont {L.}~\bibnamefont
  {Crippa}}, \bibinfo {author} {\bibfnamefont {A.}~\bibnamefont {Amaricci}},
  \bibinfo {author} {\bibfnamefont {S.}~\bibnamefont {Adler}}, \bibinfo
  {author} {\bibfnamefont {G.}~\bibnamefont {Sangiovanni}},\ and\ \bibinfo
  {author} {\bibfnamefont {M.}~\bibnamefont {Capone}},\ }\bibfield  {title}
  {\bibinfo {title} {Local versus nonlocal correlation effects in interacting
  quantum spin {Hall} insulators},\ }\href
  {https://doi.org/10.1103/PhysRevB.104.235117} {\bibfield  {journal} {\bibinfo
   {journal} {Phys. Rev. B}\ }\textbf {\bibinfo {volume} {104}},\ \bibinfo
  {pages} {235117} (\bibinfo {year} {2021})}\BibitemShut {NoStop}%
\bibitem [{\citenamefont {Richaud}\ \emph {et~al.}(2021)\citenamefont
  {Richaud}, \citenamefont {Ferraretto},\ and\ \citenamefont
  {Capone}}]{Richaud2021PRB}%
  \BibitemOpen
  \bibfield  {author} {\bibinfo {author} {\bibfnamefont {A.}~\bibnamefont
  {Richaud}}, \bibinfo {author} {\bibfnamefont {M.}~\bibnamefont
  {Ferraretto}},\ and\ \bibinfo {author} {\bibfnamefont {M.}~\bibnamefont
  {Capone}},\ }\bibfield  {title} {\bibinfo {title} {Interaction-resistant
  metals in multicomponent fermi systems},\ }\href
  {https://doi.org/10.1103/PhysRevB.103.205132} {\bibfield  {journal} {\bibinfo
   {journal} {Phys. Rev. B}\ }\textbf {\bibinfo {volume} {103}},\ \bibinfo
  {pages} {205132} (\bibinfo {year} {2021})}\BibitemShut {NoStop}%
\bibitem [{\citenamefont {Tusi}\ \emph {et~al.}(2022)\citenamefont {Tusi},
  \citenamefont {Franchi}, \citenamefont {Livi}, \citenamefont {Baumann},
  \citenamefont {Benedicto~Orenes}, \citenamefont {Del~Re}, \citenamefont
  {Barfknecht}, \citenamefont {Zhou}, \citenamefont {Inguscio}, \citenamefont
  {Cappellini}, \citenamefont {Capone}, \citenamefont {Catani},\ and\
  \citenamefont {Fallani}}]{Tusi2022NatPhys}%
  \BibitemOpen
  \bibfield  {author} {\bibinfo {author} {\bibfnamefont {D.}~\bibnamefont
  {Tusi}}, \bibinfo {author} {\bibfnamefont {L.}~\bibnamefont {Franchi}},
  \bibinfo {author} {\bibfnamefont {L.~F.}\ \bibnamefont {Livi}}, \bibinfo
  {author} {\bibfnamefont {K.}~\bibnamefont {Baumann}}, \bibinfo {author}
  {\bibfnamefont {D.}~\bibnamefont {Benedicto~Orenes}}, \bibinfo {author}
  {\bibfnamefont {L.}~\bibnamefont {Del~Re}}, \bibinfo {author} {\bibfnamefont
  {R.~E.}\ \bibnamefont {Barfknecht}}, \bibinfo {author} {\bibfnamefont
  {T.-W.}\ \bibnamefont {Zhou}}, \bibinfo {author} {\bibfnamefont
  {M.}~\bibnamefont {Inguscio}}, \bibinfo {author} {\bibfnamefont
  {G.}~\bibnamefont {Cappellini}}, \bibinfo {author} {\bibfnamefont
  {M.}~\bibnamefont {Capone}}, \bibinfo {author} {\bibfnamefont
  {J.}~\bibnamefont {Catani}},\ and\ \bibinfo {author} {\bibfnamefont
  {L.}~\bibnamefont {Fallani}},\ }\bibfield  {title} {\bibinfo {title}
  {Flavour-selective localization in interacting lattice fermions},\ }\href
  {https://doi.org/10.1038/s41567-022-01726-5} {\bibfield  {journal} {\bibinfo
  {journal} {Nature Physics}\ }\textbf {\bibinfo {volume} {18}},\ \bibinfo
  {pages} {1201} (\bibinfo {year} {2022})}\BibitemShut {NoStop}%
\bibitem [{\citenamefont {Peres}(1996)}]{Peres96}%
  \BibitemOpen
  \bibfield  {author} {\bibinfo {author} {\bibfnamefont {A.}~\bibnamefont
  {Peres}},\ }\bibfield  {title} {\bibinfo {title} {Separability criterion for
  density matrices},\ }\href {https://doi.org/10.1103/PhysRevLett.77.1413}
  {\bibfield  {journal} {\bibinfo  {journal} {Phys. Rev. Lett.}\ }\textbf
  {\bibinfo {volume} {77}},\ \bibinfo {pages} {1413} (\bibinfo {year}
  {1996})}\BibitemShut {NoStop}%
\bibitem [{\citenamefont {Horodecki}(1997)}]{Horodecki97}%
  \BibitemOpen
  \bibfield  {author} {\bibinfo {author} {\bibfnamefont {P.}~\bibnamefont
  {Horodecki}},\ }\bibfield  {title} {\bibinfo {title} {Separability criterion
  and inseparable mixed states with positive partial transposition},\ }\href
  {https://doi.org/https://doi.org/10.1016/S0375-9601(97)00416-7} {\bibfield
  {journal} {\bibinfo  {journal} {Physics Letters A}\ }\textbf {\bibinfo
  {volume} {232}},\ \bibinfo {pages} {333} (\bibinfo {year}
  {1997})}\BibitemShut {NoStop}%
\bibitem [{\citenamefont {Horodecki}\ \emph {et~al.}(2009)\citenamefont
  {Horodecki}, \citenamefont {Horodecki}, \citenamefont {Horodecki},\ and\
  \citenamefont {Horodecki}}]{HHHH_RMP}%
  \BibitemOpen
  \bibfield  {author} {\bibinfo {author} {\bibfnamefont {R.}~\bibnamefont
  {Horodecki}}, \bibinfo {author} {\bibfnamefont {P.}~\bibnamefont
  {Horodecki}}, \bibinfo {author} {\bibfnamefont {M.}~\bibnamefont
  {Horodecki}},\ and\ \bibinfo {author} {\bibfnamefont {K.}~\bibnamefont
  {Horodecki}},\ }\bibfield  {title} {\bibinfo {title} {Quantum entanglement},\
  }\href {https://doi.org/10.1103/RevModPhys.81.865} {\bibfield  {journal}
  {\bibinfo  {journal} {Rev. Mod. Phys.}\ }\textbf {\bibinfo {volume} {81}},\
  \bibinfo {pages} {865} (\bibinfo {year} {2009})}\BibitemShut {NoStop}%
\end{thebibliography}%

\end{document}